\documentclass[aps,twocolumn]{revtex4}%
\usepackage{amsfonts}
\usepackage{amsmath}
\usepackage{amssymb}
\usepackage{graphicx}%
\begin{document}
\title{Cavity Opto-Mechanics}
\author{T.J. Kippenberg}
\email{tjk@mpq.mpg.de}
\address{Max Planck Institut f\"ur Quantenoptik, Garching, Germany}
\author{and K.J. Vahala$^{2}$}
\email{vahala@caltech.edu}
\address{California Institute of Technology, Pasadena, USA}

\begin{abstract}
The coupling of mechanical and optical degrees of freedom via radiation
pressure has been a subject of early research in the context of gravitational
wave detection. Recent experimental advances have allowed studying for the
first time the modifications of mechanical dynamics provided by radiation
pressure. This paper reviews the consequences of back-action of light confined
in whispering-gallery dielectric micro-cavities, and presents a unified
treatment of its two manifestations: notably the parametric instability
(parametric amplification) and radiation pressure back-action cooling.
Parametric instability offers a novel "photonic clock" which is driven purely
by the pressure of light. In contrast, radiation pressure cooling can surpass
existing cryogenic technologies and offers cooling to phonon occupancies below
unity and provides a route towards cavity Quantum Optomechanics

\end{abstract}
\maketitle

\section{Introduction}

Recent years have witnessed a series of developments at the intersection of
two, previously distinct subjects. Optical (micro-)cavities\cite{Vahala2003}
and micro (nano) mechanical resonators\cite{Craighead2000}, each a subject in
their own right with a rich scientific and technological history, have, in a
sense, become entangled experimentally by the underlying mechanism of optical,
radiation pressure forces. These forces and their related physics have been of
major interest in the field of atomic
physics\cite{Hansch1975,WinelandDrullinger1978,ChuHollberg1985,Stenholm1986}
for over 5 decades and the emerging opto-mechanical context has many parallels
with this field.

The opto-mechanical coupling between a moving mirror and the radiation
pressure of light has first appeared in the context of interferometric
gravitational wave experiments. Owing to the discrete nature of photons, the
quantum fluctuations of the radiation pressure forces give rise to the so
called \textit{standard quantum limit}%
\cite{Caves1981,JacobsTittonen1999,TittonenBreitenbach1999}. In
addition to this \textquotedblleft\textit{quantum
back-action}\textquotedblright\ effect, the pioneering work of V.
Braginsky\cite{Braginsky1977} predicted that the radiation pressure
of light, confined within an interferometer (or resonator), gives
rise to the effect of dynamic back-action (which is a classical
effect, in contrast to the aforementioned quantum back-action) owing
to the finite cavity decay time. The resulting phenomena, which are
the parametric instability (and associated mechanical oscillation)
and opto-mechanical back-action cooling represent\textit{\ two sides
of the same underlying \textquotedblleft dynamic
back-action\textquotedblright\ mechanism}. Later, theoretical work
has proposed to using the radiation-pressure coupling to realize
quantum non-demolition measurements of the light
field\cite{BraginskyKhalili1992} and as a means to create
non-classical states of the light field\cite{ManciniTombesi1994} and
the mechanical system\cite{BoseJacobs1997}. It is noted that the
effect of dynamic back-action is of rather general relevance and can
occur outside of the opto-mechanical context. Indeed, the same
dynamic back-action phenomena have been predicted to occur in
systems where a mechanical oscillator is parametrically coupled to
an electromagnetic resonant system, and have indeed been observed in
an electronic resonance of a superconducting single electron
transistor coupled to a mechanical
oscillator\cite{LaHayeBuu2004,NaikBuu2006} (or LC
circuit\cite{Braginsky1977,BrownBritton2007}).

\begin{figure}[ptbh]
\centering\includegraphics[width=7.6cm]{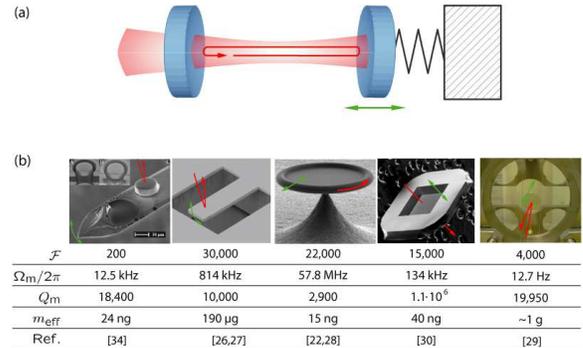}\caption{(a)A cavity
optomechanical system consisting of a Fabry Perot cavity with a harmonically
bound end mirror. Panel (b): Different physical realizations of cavity
optomechanical experiments employing cantilevers, micro-mirrors,
micro-cavities, nano-membranes and macroscopic mirror modes. Red and green
arrows represent the optical trajectory and mechanical motion.}%
\label{Figure1}%
\end{figure}\qquad

The experimental manifestations of optomechanical coupling \ by radiation
pressure have been observable for some time. For instance, radiation pressure
forces were observed in the pioneering work of H. Walther at the MPQ,
manifesting themselves in radiation pressure (pondermotive)
bistability\cite{DorselMcCullen1983}, while even earlier work in the microwave
domain had been carried out by Braginsky\cite{Braginsky1977}. Moreover the
modification of mechanical oscillator stiffness caused by radiation pressure,
the \textquotedblleft\textit{optical spring}\textquotedblright,\ have also
been observed\cite{SheardGray2004}. However, in contrast to these
\textit{static} manifestations of radiation pressure, the \textit{dynamic}
manifestations of radiation pressure forces on micro- and nano-mechanical
objects have only recently become an experimental reality. Curiously, while
the theory of dynamic back action was motivated by consideration of precision
measurement in the context of gravitational wave detection using large
interferometers\cite{BraginskyStrigin2001}, the first observation of this
mechanism was reported in 2005 at a vastly different size scale in microtoroid
cavities\cite{KippenbergRokhsari2005,
RokhsariKippenberg2005,CarmonRokhsari2005}. These observations were focused on
the radiation pressure induced parametric
instability\cite{BraginskyStrigin2001}. Subsequently, the reverse
mechanism\cite{BraginskyVyatchanin2002} (back-action cooling) has been
exploited to cool
cantilevers\cite{ArcizetCohadon2006Nature,GiganBohm2006,PoggioDegen2007},
microtoroids\cite{SchliesserDelHaye2006} and macroscopic mirror
modes\cite{CohadonHeidmann1999, CorbittChen2007} as well as mechanical
nano-membranes\cite{ThompsonZwickl2007}. We note that this technique is
different than the earlier demonstrated radiation-pressure feedback
cooling\cite{ManciniVitali1998,CohadonHeidmann1999}, which uses electronic
feedback analogous to \textquotedblleft Stochastic Cooling\textquotedblright%
\cite{Vandermeer1985} of ions in storage rings and which can also provide very
efficient cooling as demonstrated in recent
experiments\cite{KlecknerBouwmeester2006,PoggioDegen2007,ArcizetCohadon2006}.
Indeed, research in this subject has experienced a remarkable acceleration
over the past three years as researchers in diverse fields such as optical
microcavities\cite{Vahala2003}, micro and nano-mechanical
resonators\cite{Craighead2000} and quantum optics pursue a common set of
scientific goals set forward by a decade-old theoretical framework. Indeed,
there exists a rich theoretical history that considers the implications of
optical forces in this new context. Subjects ranging from
entanglement\cite{GiovannettiMancini2001, ManciniGiovannetti2002}; generation
of squeezed states of light\cite{ManciniTombesi1994}; to measurements at or
beyond the standard quantum
limit\cite{BraginskyKhalili1992,CourtyHeidmann2003,ArcizetBriant2006}; and
even to tests of quantum theory itself are in play
here\cite{MarshallSimon2003}. On the practical side, there are opportunities
to harness these forces for new metrology tools\cite{ArcizetCohadon2006} and
even for new functions on a semiconductor chip (e.g.,
oscillators\cite{KippenbergRokhsari2005, RokhsariKippenberg2005}, optical
mixers\cite{HosseinZadehVahala2007_2}, and tuneable optical filters and
switches\cite{PovinelliJohnson2005,EichenfeldMichael2007}. It seems clear that
a new field of \textit{cavity optomechanics} has emerged, and will soon evolve
into \textit{cavity quantum optomechanics} (cavity QOM) whose goal is the
observation and exploration of quantum phenomena of mechanical
systems\cite{SchwabRoukes2005} as well as quantum phenomena involving both
photons and mechanical systems.
The realization of dynamical, opto-mechanical coupling in which radiation
forces mediate the interaction, is a natural outcome of underlying
improvements in the technologies of optical (micro) cavities and mechanical
micro (nano-) resonators. Reduction of loss (increasing optical and mechanical
Q) and reductions in form factor (modal volume) have enabled a regime of
operation in which optical forces are
dominant\cite{KippenbergRokhsari2005,RokhsariKippenberg2005,
ArcizetCohadon2006, GiganBohm2006, KlecknerBouwmeester2006,
SchliesserDelHaye2006, CorbittChen2007, CorbittChen2007,
EichenfeldMichael2007, ThompsonZwickl2007}. This coupling also requires
coexistence of both high-Q optical and high-Q mechanical modes. Such
coexistence has been achieved in the geometries illustrated in Figure
\ref{Figure1}. It also seems likely that other optical microcavity geometries
such as photonic crystals\cite{AkahaneAsano2003} can exhibit the dynamic
back-action effect provided that structures are modified to support high-Q
mechanical modes.

To understand how the coupling of optical and mechanical degrees of freedom
occurs in any of the depicted geometries, one need only consider the schematic
in the upper panel of Figure \ref{Figure1}. Here, a Fabry-Perot optical cavity
is assumed to feature a mirror that also functions as a mass-on-a-spring (i.e.
is harmonically suspended). Such a configuration can indeed be encountered in
gravitational wave laser interferometers (such as LIGO) and is also, in fact,
a direct representation of the \textquotedblleft cantilever
mirror\textquotedblright\ embodiment in the lower panel within Figure
\ref{Figure1}. In addition it is functionally equivalent to the case of a
microtoroid embodiment (also shown in the lower panel), where the toroid
itself provides both the optical modes as well as mechanical breathing modes
(see Figure \ref{Figure1} and discussion below). Returning to the upper panel,
incident optical power that is resonant with a cavity mode creates a large
circulating power within the cavity. This circulating power exerts a force
upon the \textquotedblleft mass-spring\textquotedblright\ mirror, thereby
causing it to move. Reciprocally, the mirror motion results in a new optical
round trip condition, which modifies the detuning of the cavity resonance with
respect to the incident field. This will cause the system to simply establish
a new, static equilibrium condition. The nonlinear nature of the coupling in
such a case can manifest itself as a hysteretic behavior and was observed over
two decades ago in the work by Walther et. al\cite{DorselMcCullen1983}.
However, if the mechanical and optical Q's are sufficiently high (as is
further detailed in what follows, such that the mechanical oscillation period
is comparable or exceeds the cavity photon lifetime) a new set of dynamical
phenomena can emerge, related to mechanical amplification and cooling.

In this paper, we will give the first unified treatment of this subject.
Although microtoroid optical microcavities will be used as an illustrative
platform, the treatment and phenomena are universal and pertain to any cavity
opto-mechanical systems supporting high Q optical and mechanical modes. In
what follows we begin with a theoretical framework through which dynamic
back-action can be understood. The observation of micromechanical oscillation
will then be considered by reviewing both old and new experimental results
that illustrate the basic phenomena\cite{KippenbergRokhsari2005,
RokhsariKippenberg2005,CarmonRokhsari2005}. Although this mechanism has been
referred to as the parametric instability, we show that it is more properly
defined in the context of mechanical amplification and regenerative
oscillation. For this reason, we introduce and define a mechanical gain, its
spectrum, and, correspondingly, a threshold for oscillation. Mechanical
cooling is introduced as the reverse mechanism to amplification. We will then
review the experimental observation of cooling by dynamic
back-action\cite{SchliesserDelHaye2006,SchliesserRiviere2007} and also the
quantum limits of cooling using back
action\cite{MarquardtChen2007,WilsonRaeNooshi2007}.

Finally, we will attempt to discuss some of the possible future directions for
this new field of research.

\section{Theoretical Framework of Dynamic Back-action}
\subsection{Coupled Mode Equations}
Dynamic back-action is a modification of mechanical dynamics caused by
non-adiabatic response of the optical field to changes in the cavity size. It
can be understood through the coupled equations of motion for optical and
mechanical modes, which can be derived from a single Hamiltonian\cite{Law1995}.%

\begin{equation}
\frac{da}{dt}=i\Delta(x)a-\left(  \frac{1}{2\tau_{0}}+\frac{1}{2\tau_{ex}%
}\right)  a+i\sqrt{\frac{1}{\tau_{ex}}}\label{Equation1}%
\end{equation}%
\begin{equation}
\frac{d^{2}x}{dt^{2}}+\frac{\Omega_{m}}{2Q_{m}}\frac{dx}{dt}+\Omega_{m}%
^{2}x=\frac{F_{RP}(t)}{m_{eff}}+\frac{F_{L}(t)}{m_{eff}}=\frac{\zeta}%
{cm_{eff}}\frac{|a|^{2}}{T_{rt}}+\frac{F_{L}(t)}{m_{eff}}\label{Equation2}%
\end{equation}

Aside from the $x-$dependence, the first equation governs the dynamics of the
optical field according to the formalism of H. Haus\cite{Haus1989}, i.e.
$|a|^{2}$ is the stored cavity energy, whereas $|s|^{2}$ denotes the launched
input power upon the cavity system. Moreover the optical field decays with a
rate $\frac{1}{2\tau}=\frac{1}{2\tau_{0}}+\frac{1}{2\tau_{ex}}$and
$\Delta(x)=$ $\omega-\omega_{0}(x)$ accounts for the detuning of the pump
laser frequency $\omega$ with respect to the cavity resonance $\omega_{0}(x)$
(which, as shown below, depends on the mechanical coordinate, $x$). The power
coupling rate into outgoing modes is described by the rate $1/\tau_{ex}$,
whereas the intrinsic cavity loss rate is given by $1/\tau_{0}$. In the
subsequent discussion, the photon decay rate is also used $\kappa\equiv1/\tau$.

The second equation describes the mechanical coordinate ($x$) accounting for
the movable cavity boundary (i.e. mirror), which is assumed to be harmonically
bound and undergoing harmonic oscillation at frequency $\Omega_{m}$ with a
power dissipation rate $\Gamma_{m}=$ $\frac{\Omega_{m}}{Q_{m}}.$ Moreover,
$m_{eff}$ is the effective mass of the mirror mode and will be discussed in a
later section. This mass describes in large part the strength of the coupling
between optical and mechanical mode. For an excellent treatment of its
determination and derivation the reader is referred to
reference\cite{PinardHadjar1999}. The radiation pressure forcing function is
given by $F_{RP}(t)$ $=\frac{\zeta}{c}\frac{|a|^{2}}{T_{rt}}$, where the
dimensionless parameter $\zeta$ takes on the value $2\pi n$ for a
whispering-gallery-mode micro-cavity (consisting of a dielectric material with
refractive index $n$), and $\zeta=2$ for a Fabry Perot cavity; and where
$T_{rt}$ is the cavity round trip time (note that the intracavity circulating
power is given by $\frac{|a|^{2}}{T_{rt}}$). Moreover the term $F_{L}(t)$
denotes the random Langevin force, and obeys $\left\langle F_{L}%
(t)F_{L}(t^{\prime})\right\rangle =\Gamma_{m}k_{B}T_{R}m_{eff}\delta
(t-t^{\prime})$, where $k_{B}$ is the Boltzmann constant and $T_{R}$ is the
temperature of the reservoir. The Langevin force ensures that the fluctuation
dissipation theorem is satisfied, such that the total steady state energy in
the (classical) mechanical mode $E_{m}$ (in the absence of laser radiation) is
given by $E_{m}=\int\Omega^{2}m_{eff}\left\vert \chi(\Omega)F_{L}%
(\Omega)\right\vert ^{2}d\Omega=k_{B}T_{R}$, where the mechanical
susceptibility $\chi(\Omega)=m_{eff}^{-1}/(i\Omega\Gamma_{m}+\Omega_{m}%
^{2}-\Omega^{2})$ has been introduced. Of special interest in the first
equation is the optical detuning $\Delta(x)$ which provides coupling to the
second equation through the relation:%

\begin{equation}
\Delta(x)=\Delta+\frac{\omega_{0}}{R}x \label{Eq3}%
\end{equation}

This relation assumes that, under circumstances in which the mass-spring is at
rest $(x=0)$, the optical pump is detuned by $\Delta$ relative to the optical
mode resonance. Two cases of interest, both illustrated in Figure
\ref{Figure1}, will emerge: blue detuned ($\Delta>0$) and red-detuned
($\Delta<0$) operation of the pump-wave relative to the cavity resonance. It
is important to note that quadratic coupling can also be realized, e.g.
Ref\cite{ThompsonZwickl2007}, however, this case will not be considered here.

Before discussing the physics associated with the optical delay (which gives
rise to dynamical back-action), we briefly divert to a \textit{static}
phenomena that is associated with the steady state solutions of the above
coupled equations: the mirror bistability\cite{DorselMcCullen1983} and
multi-stability\cite{MarquardtHarris2006}. Indeed, as noted earlier, the
radiation-pressure-induced bi-stability was observed over two decades
ago\cite{DorselMcCullen1983} using a harmonically-suspended mirror. In brief,
considering purely the static solutions for the displacement (i.e.
$\overline{x}$) of the above set of equations, it becomes directly evident
that the equilibrium position of the mechanical oscillator will depend upon
the intra-cavity power. Since the latter is again coupled to the mechanical
displacement, this leads to a cubic equation for the mirror position
$\overline{x}$ as a function of applied power:
\begin{equation}
\frac{\tau^{2}}{\tau_{ex}}|s|^{2}=\frac{cm_{eff}}{\zeta}\Omega_{m}%
^{2}\overline{x}\left(  4\tau^{2}\left(  \Delta+\frac{\omega_{0}}{R}%
\overline{x}\right)  ^{2}+1\right)  \label{Eq4}%
\end{equation}

For sufficiently high power, this leads to bi-stable behavior
(namely for a given detuning and input power the mechanical position
can take on several possible values).
\subsection{Modifications due to Dynamic Back-action: Method of Retardation
Expansion}
 The circulating optical power will vary in response to
changes in the coordinate $"x"$. The delineation of this response
into adiabatic and non-adiabatic contributions provides a starting
point to understand the origin of dynamic back-action and its two
manifestations. This delineation is possible by formally integrating
the above equation for the circulating field, and treating the term
$x$ wihin it as a perturbation. Furthermore, if the time variation
of $x$ is assumed to be slow on the time scale of the optical cavity
decay time (or equivalently if $\kappa\gg\Omega_{m}$) then an
expansion into orders of retardation is possible. Keeping only terms
up to $\frac{dx}{dt}$ in the series expansion, the circulating
optical power (and hence the forcing
function in mechanical oscillator equation) can be expressed as follows,%

\begin{multline}
P_{cav}=P_{cav}^{0}+P_{cav}^{0}\left(  \frac{8\Delta\tau^{2}}{4\tau^{2}%
\Delta^{2}+1}\right)  \frac{\omega_{0}}{R}x\label{Eq6}\\
-\tau P_{cav}^{0}\left(  \frac{8\Delta\tau^{2}}{4\tau^{2}\Delta^{2}+1}\right)
\frac{\omega_{0}}{R}\frac{dx}{dt}%
\end{multline}

Here the power $P_{cav}^{0}=\frac{|a|^{2}}{T_{rt}}=\frac{\tau}{T_{rt}}%
\frac{2\tau/\tau_{ex}}{4\Delta^{2}\tau^{2}+1}|s|^{2}$ denotes the power in the
cavity for zero mechanical displacement. The circulating power can also be
written as $P_{cav}^{0}=\mathcal{F}/\pi\cdot C\cdot|s|^{2}$, where
$C=\frac{\tau/\tau_{ex}}{4\Delta^{2}\tau^{2}+1}$ and where $\mathcal{F}%
=2\pi\frac{\tau}{T_{rt}}$ denotes the cavity Finesse. The first two terms in
this series provide the adiabatic response of the optical power to changes in
position of the mirror. Intuitively, they give the instantaneous variations in
coupled power that result as the cavity is \textquotedblleft
tuned\textquotedblright\ by changes in $``x\textquotedblright$. It is apparent
that the $x-$dependent contribution to this adiabatic response (when input to
the mechanical-oscillator equation-of-motion through the forcing function
term) provides an optical-contribution to the stiffness of the mass-spring
system. This so-called \textit{\textquotedblleft optical
spring\textquotedblright} effect (or\textit{ "light induced rigidity"}) has
been observed in the context of LIGO\cite{SheardGray2004} and also in
microcavities\cite{HosseinZadehVahala2007}. The corresponding change in spring
constant leads to a frequency shift, relative to the unpumped mechanical
oscillator eigenfrequency, as given by (where $P=|s|^{2}$ denotes the input power):%
\begin{equation}
\Delta\Omega_{m}\underset{\kappa\gg\Omega_{m}}{=}\mathcal{F}^{2}\frac
{8n^{2}\omega_{0}}{\Omega_{\mathrm{m}}mc^{2}}C\cdot\left[  \frac{2\Delta\tau
}{\left(  4\tau^{2}\Delta^{2}+1\right)  }\right]  P \label{Eq7}%
\end{equation}
The non-adiabatic contribution in equation\ref{Eq6} is proportional to the
velocity of the mass-spring system. When input to the mechanical-oscillator
equation, the coefficient of this term is paired with the intrinsic mechanical
damping term and leads to the following damping rate given by (for a
whispering gallery mode cavity of radius $R$):%

\begin{equation}
\Gamma\underset{\kappa\gg\Omega_{m}}{=}-\mathcal{F}^{3}\frac{8n^{3}\omega
_{0}R}{\Omega_{\mathrm{m}}mc^{3}}C\cdot\left[  \frac{8\Delta\tau}{\left(
4\tau^{2}\Delta^{2}+1\right)  ^{2}}\right]  P \label{Eq8}%
\end{equation}

Consequently, the modified (effective) damping rate of the mechanical
oscillator is given by:%
\begin{equation}
\Gamma_{eff}=\Gamma+\Gamma_{m}%
\end{equation}

In both equation \ref{Eq7} and \ref{Eq8}we have stressed the fact that these
expressions are valid only in the weak retardation regime in which $\kappa
\gg\Omega_{m}$. The sign of $\Gamma$ (and the corresponding direction of power
flow) depends upon the relative detuning of the optical pump with respect to
the cavity resonant frequency. In particular, a red-detuned pump ( $\Delta<0$)
results in a sign such that optical forces augment intrinsic mechanical
damping, while a blue-detuned pump ( $\Delta>0$) reverses the sign so that
damping is offset (negative damping or amplification). It is important to note
that the cooling rate in the weak-retardation regime depends strongly
($\propto\mathcal{F}^{3}$) on the optical finesse, which has been
experimentally verified as discussed in section 4.2. Note also that maximum
cooling or amplification rate for given power occurs when the laser is detuned
to the maximum slope of the cavity lorenzian; these two cases are illustrated
in Figure \ref{Figure1}. These modifications have been first derived by
Braginsky\cite{Braginsky1977} more than 3 decades ago and are termed
\textit{dynamic back-action}. Specifically, an optical probe used to ascertain
the position of a mirror within an optical resonator, will have the
side-effect of altering the dynamical properties of the mirror (viewed as a
mass-spring system).

\begin{figure}[ptb]
\centering\includegraphics[width=7.6cm]{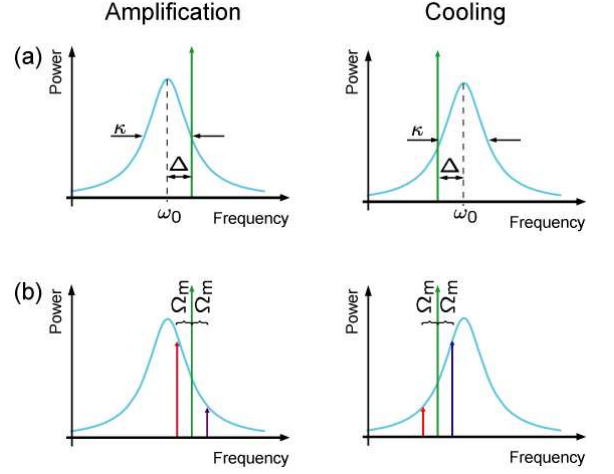}\caption{The two
manifestations of dynamic back-action: blue-detuned and red-detuned pump wave
(green) with respect to optical mode line-shape (blue) provide mechanical
amplification and cooling, respectively. Also shown in the lower panels are
motional sidebands (Stokes and anti-Stokes fields) generated by mirror
vibration and subsequent Doppler-shifts of the circulating pump field. The
amplitudes of these motional sidebands are asymmetric owing to cavity
enhancement of the Doppler scattering process.}%
\label{Figure2}%
\end{figure}

\begin{figure}[ptb]
\centering\includegraphics[width=7.6cm]{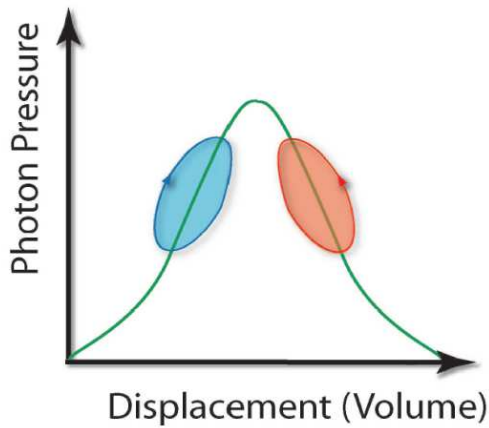}\caption{Work done during
one cycle of mechanical oscillation can be understood using a PV diagram for
the radiation pressure applied to a piston-mirror versus the mode volume
displaced during the cycle. In this diagram the cycle follows a contour that
circumscribes an area in PV space and hence work is performed during the
cycle. The sense in which the contour is traversed (clockwise or
counterclockwise) depends upon whether the pump is blue or red detuned with
respect to the optical mode. Positive work (amplification) or negative work
(cooling) are performed by the photon gas on the piston mirror in the
corresponding cases.}%
\label{Figure3}%
\end{figure}

The direction of power-flow is also determined by the sign of the pump
frequency detuning relative to cavity resonant frequency. Damping (red tuning
of the pump) is accompanied by power flow from the mechanical mode to the
optical mode. This flow results in cooling of the mechanical mode.
Amplification (blue tuning of pump) is, not surprisingly, accompanied by net
power flow from the optical mode to the mechanical mode. This case has also
been referred to as \textit{heating}, however, it is more appropriately
referred to as \textit{amplification} since the power flow in this direction
performs work on the mechanical mode. The nature of power flow between the
mechanical and optical components of the system will be explored here in
several ways, however, one form of analysis makes contact with the
thermodynamic analogy of cycles in a Clapeyron or Watt diagram (i.e. a
pressure-volume diagram). In the present case -- assuming the mechanical
oscillation period to be comparable to or longer than the cavity lifetime --
such a diagram can be constructed to analyze power flow resulting from cycles
of the coherent radiation gas interacting with a movable
piston-mirror\cite{Karrai2006} (see Figure \ref{Figure3}). In particular, a
plot of radiation pressure exerted on the piston-mirror versus changes in
optical mode volume provides a coordinate space in which it is possible to
understand the origin and sign of work done during one oscillation cycle of
the piston mirror. Considering the oscillatory motion of the piston-mirror at
some eigenfrequency $\Omega_{m}$, then because pressure (proportional to
circulating optical power) and displaced volume (proportional to $x$) have a
quadrature relationship (through the dynamic back-action term involving the
velocity $\frac{dx}{dt}$), the contour for a PV cycle will encompass a
non-zero area, giving the net-work performed during one cycle of mechanical
oscillation. Note that the area reduces as the photon lifetime shortens (i.e.,
as retardation is weakened). Also, the sense in which the PV cycle is
traversed is opposite for the two cases of red and blue detuning of the pump
(i.e., the area and hence work-done changes sign). For blue detuning, the
radiation gas does net work on the piston while the reverse is true for red
detuning. The fact that cooling is possible for the case of a red detuned pump
is the result of both: the sign of work in this case being such that the
piston does positive work on the photon gas, and, equally important, that the
photon gas (if in a coherent state) provides only quantum, back-action on the
piston. This makes the photon gas effectively very cold.

The cooling and amplification processes can also be understood in terms of the
creation of stokes and anti-Stokes sidebands\cite{KippenbergRokhsari2005}.
Oscillatory motion of the cavity mirror will create side bands on the probe
wave as the circulating optical power is Doppler-shifted by the mirror's
motion (or equivalently the expansion and contraction of the whispering
gallery in the case of a toroid or disk resonator). As shown in Figure
\ref{Figure2}, these motional sidebands will have asymmetric amplitudes owing
to the fact that the pump wave frequency is detuned from the cavity resonance.
The two cases of interest (red and blue detuning) are illustrated in the
Figure \ref{Figure2} and produce opposite asymmetry. Intuitively, the sideband
that is closer to the cavity resonance has its amplitude resonantly enhanced.
This will be rigorously derived in the next section. This asymmetry indicates
a net deficit (blue detuned) or surplus (red detuned) of power in the
transmitted pump wave. The sideband asymmetry also produces strong amplitude
modulation that can be measured as a photocurrent upon detection of the
transmitted power from the resonator.

Amplification and damping (i.e. cooling) have been verified experimentally. In
particular, a mechanical mode subjected to dynamic backaction through a
properly detuned optical pump wave will exhibit a thermal noise spectrum whose
line-shape is modified by the presence of added optical damping or gain
(negative damping). In either case, the damping rate will depend in a linear
fashion on the coupled optical pump power for fixed pump detuning. Before
proceeding with the further development of the theory, it is useful to
consider the measurement of this damping dependence in an actual system
(Further details regarding the realization of amplification and cooling will
be the subject of section 4 and 5). Measurement of this behavior is possible
by probing the motion of the cavity mirror using the transmitted optical pump
wave. If the pump wave is detuned relative to the cavity resonance, it is
power-modulated by the resonator since the mechanical motions vary in time the
cavity resonance. (Equivalently, the asymmetric motional sidebands on the
transmitted pump are detected.) The detected amplitude modulation contains
information about the underlying mechanical motion. If, for example, the
mirror is undergoing regenerative oscillations (see section 4 below), then
spectral analysis of the photocurrent will reveal the oscillation frequency
(and even the effective temperature) and measurement of the modulation power
can be used to ascertain the mechanical oscillation amplitude. In cases where
the mirror is excited only by thermal energy, the spectrally-broad thermal
excitation (upon measurement as a photocurrent spectral density) provides a
way to observe the oscillator lineshape and thereby determine its linewidth
and effective damping rate. If the optical pump wave is weak, this damping
rate will reflect the intrinsic loss (intrinsic mechanical $Q_{m}=\Omega
_{m}/\Gamma_{m}$) of the mechanical mode. On the other hand, as the probe
power is increased, it will modify the damping rate causing the line to narrow
(blue detuned) or broaden (red detuned) in accordance with the above model.

\begin{figure}[ptbh]
\centering\includegraphics[width=7.6cm]{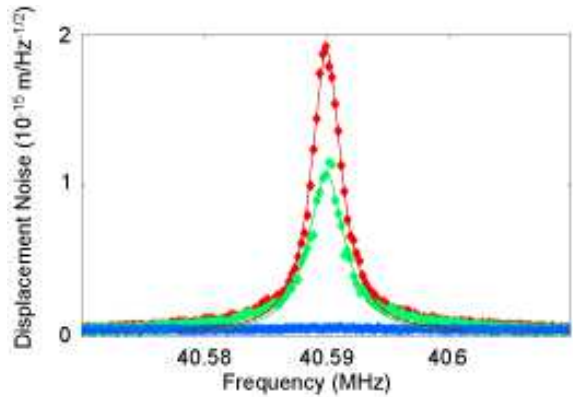}\caption{Dynamics in the
weak retardation regime. Experimental displacement spectral density functions
for a mechanical mode with eigenfrequency 40.6 MHz measured using three,
distinct pump powers for both blue and red pump detuning. The mode is
thermally excited (green data) and its linewidth can be seen to narrow under
blue pump detuning (red data) on account of the presence of mechanical gain
(not sufficient in the present measurement to excite full, regenerative
oscillations); and to broaden under red pump detuning on account of radiation
pressure damping (blue data). }%
\label{Figure4}%
\end{figure}

\begin{figure}[ptbh]
\centering\includegraphics[width=7.6cm]{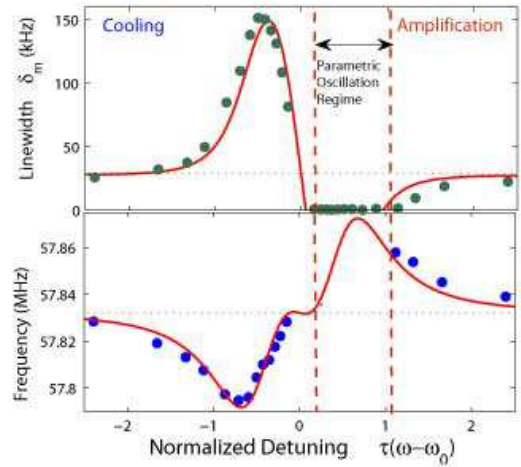}\caption{Upper panel shows
mechanical linewidth ($\delta_{m}=\Gamma_{eff}/2\pi$) and shift in mechanical
frequency (lower panel) measured versus pump wave detuning in the regime where
$\Omega_{m}\ll\kappa$. For negative (positive) detuning cooling
(amplification) occurs. The region between the dashed lines denotes the onset
of the parametric oscillation, where gain compensates mechanical loss. Figure
stems from reference\cite{SchliesserDelHaye2006}. Solid curves are theoretical
predictions based on the sideband model (see section 2.2). }%
\label{Figure5}%
\end{figure}

Figure \ref{Figure4} presents such lineshape data taken using a microtoroid
resonator. The power spectral density of the photocurrent is measured for a
mechanical mode at an eigenfrequency of 40.6 $%
\operatorname{MHz}%
$; three spectra are shown, corresponding to room temperature intrinsic motion
(i.e. negligible pumping), mechanical amplification and cooling. In addition
to measurements of amplification and damping as a function of pump power (for
fixed detuning), the dependence of these quantities on pump detuning (with
pump power fixed) can also be measured\cite{SchliesserDelHaye2006}.
Furthermore, pulling of the mechanical eigenfrequency (caused by the radiation
pressure modification to mechanical stiffness) can also be
studied\cite{HosseinZadehVahala2007}. A summary of such data measured using a
microtoroid in the regime where $\kappa\gtrsim\Omega_{m}$ is provided in
Figure \ref{Figure5}. Both the case of red- (cooling) and blue-
(amplification) pump detuning are shown. Furthermore, it can be seen that pump
power was sufficient to drive the mechanical system into regenerative
oscillation over a portion of the blue detuning region (section of plot where
linewidth is nearly zero). For comparison, the theoretical prediction is shown
as the solid curve in the plots. Concerning radiation-pressure-induced
stiffness, it should be noted that for red-detuning, the frequency is pulled
to lower frequencies (stiffness is reduced) while for blue-detuning the
stiffness increases and the mechanical eigenfrequency shifts to higher values.
This is in dramatic contrast to similar changes that will be discussed in the
next section. While in Figure \ref{Figure5} the absolute shift in the
mechanical eigenfrequency is small compared to $\Omega_{m}$, it is interesting
to note that it is possible for this shift to be large. Specifically,
statically unstable behavior is possible if the total spring constant reaches
a negative value. This has indeed been observed experimentally in gram scale
mirrors coupled to strong intra-cavity fields by the LIGO group at
MIT\cite{CorbittChen2007}.

While the above approach provides a convenient way to understand the origin of
gain and damping and their relationship to non-adiabatic response, a more
general understanding of the dynamical behavior requires an extension of the
formalism. Cases where the mechanical frequency, itself, varies rapidly on the
time scale of the cavity lifetime cannot be described correctly using the
above model. From the viewpoint of the sideband picture mentioned above, the
modified formalism must include the regime in which the sideband spectral
separation from the pump can be comparable or larger than the cavity linewidth.

\subsection{Sideband Formalism}
It is important to realize that the derivation of the last section only
applies to the case where the condition $\kappa\gg\Omega_{m}$ is satisfied. In
contrast, a perturbative expansion of the coupled mode equations (eqns.
\ref{Equation1} and \ref{Equation2}) gives an improved description that is
also valid in the regime where the mechanical frequency is comparable to or
even exceeds the cavity decay rate (where $\Omega_{m}\gg$ $\kappa$ is the so
called \textit{resolved sideband regime}). The derivation that leads to these
results\cite{SchliesserDelHaye2006} is outlined here. For most experimental
considerations (in particular for the case of cooling) the quantity
$\varepsilon=\frac{x}{2R}$ is very small, and a perturbative expansion of the
intracavity field in powers of this parameter is possible, i.e.%
\[
a\equiv\sum_{n=0}^{\infty}\varepsilon^{n}a_{n}%
\]
Here, the zeroth order perturbation amplitude ($a_{0}$) describes the cavity
build-up factor in the absence of the opto-mechanical coupling%

\begin{equation}
a_{0}(t)=\frac{i\sqrt{\frac{1}{\tau_{ex}}}}{i\Delta-\frac{1}{2\tau}%
}se^{i\omega t} \label{E10}%
\end{equation}
where the detuning is given by $\Delta$. Making the assumption that the
mechanical system is undergoing harmonic motion $x(t)=x\cos(\Omega_{m}t),$ it
is possible to solve for the first-order perturbation term:%
\[
a_{1}(t)=a_{0}(t)2\tau\omega_{0}\left(  \frac{e^{i\Omega_{m}t}}{2i(\Delta
+\Omega_{m})\tau-1}+\frac{e^{-i\Omega_{m}t}}{2i(\Delta-\Omega_{m})\tau
-1}\right)
\]

\begin{figure}[ptbh]
\centering\includegraphics[width=7.6cm]{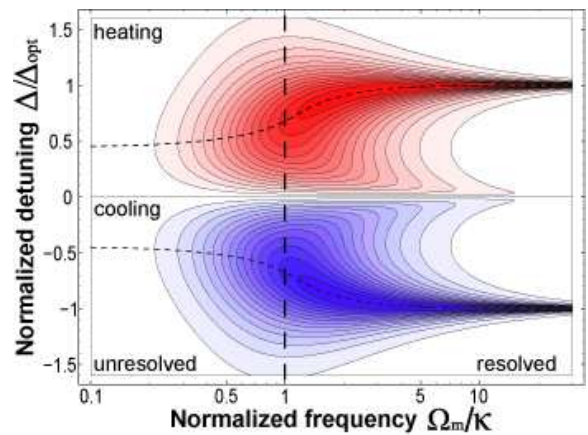}\caption{The mechanical
gain rate and cooling rate as a function of detuning and normalized mechanical
frequency. Also shown is the optimum amplification and cooling rate for fixed
frequency (dotted lines). In the simulation, pump power and cavity dimension
are fixed parameters. }%
\label{Figure7}%
\end{figure}

Inspection shows that this 1st-order term consists of two, independent fields,
a frequency upshifted anti-Stokes sideband $(\omega_{AS}=\omega+\Omega_{m})$,
and a frequency down shifted $(\omega_{S}=\omega-\Omega_{m})$ Stokes sideband
produced by the harmonic mechanical motion. These fields account for Doppler
shifting of the circulating pump field caused by the motion of the mirror or
dielectric cavity. The parameter $\varepsilon$ is sufficiently small to
neglect the higher-order terms in the perturbation. The radiation pressure
force acting on the mechanical oscillator in the toroidal (or other whispering
gallery mode) cavity is $F_{RP}(t)$ $=\frac{2\pi n}{c}\frac{|a|^{2}}{T_{rt}}%
$(and for a Fabry-Perot cavity $F_{RP}(t)$ $=\frac{2}{c}\frac{|a|^{2}}{T_{rt}%
}$). Here, $T_{rt}$ is the cavity round trip time and $\left\vert a\right\vert
^{2}$ and is to 1$^{st}$ order by $\left\vert a\right\vert ^{2}\cong\left\vert
a_{0}\right\vert ^{2}+2\varepsilon\Re{(a_{0}^{\ast}a_{1})}$. The radiation
pressure force can now be expressed in terms of \textit{in-phase} and\textit{
quadrature} components (with respect to the harmonic displacement, assumed
above):%
\[
F_{RP}(t)=\cos(\Omega_{m}t)F_{I}+\sin(\Omega_{m}t)F_{Q}%
\]
Where the in-phase component is given by:%
\begin{multline}
{\small F}_{I}{\small (t)}{\small =-\varepsilon}\frac{4\pi n}{cT_{rt}}%
\frac{\omega_{0}\tau^{2}/\tau_{ex}}{4\Delta^{2}\tau^{2}+1}\times\\
\left(  \frac{2(\Delta+\Omega_{m})}{4(\Delta+\Omega_{m})^{2}\tau^{2}+1}%
-\frac{2(\Delta-\Omega_{m})}{4(\Delta-\Omega_{m})^{2}\tau^{2}+1}\right)
{\small P}%
\end{multline}
and the quadrature component takes on the form:%
\begin{multline}
F_{Q}(t)=-\varepsilon\frac{4\pi n}{cT_{rt}}\frac{\omega_{0}\tau^{2}/\tau_{ex}%
}{4\Delta^{2}\tau^{2}+1}\times\\
\left(  \frac{2\tau}{4(\Delta+\Omega_{m})^{2}\tau^{2}+1}-\frac{2\tau}%
{4(\Delta-\Omega_{m})^{2}\tau^{2}+1}\right)  P
\end{multline}
The in-phase part of this force is responsible for changes in rigidity whereas
the quadrature part is responsible for changes in the damping factor, which
leads to cooling (or amplification). From the force, the net power $(P_{m})$
transferred from the mechanical mode to (or from) the optical mode can be
calculated via the relation $\left\langle P_{m}\right\rangle =\left\langle
F_{Q}(t)\cdot\dot{x}\right\rangle $ yielding:%
\begin{multline}
\left\langle P_{m}\right\rangle =\left\langle x^{2}\right\rangle \frac{2\pi
n}{cT_{rt}}\frac{\omega_{0}1/\tau_{ex}}{4\tau^{2}\Delta\omega^{2}+1}%
\frac{\Omega_{m}}{2R}\times\\
\left(  \frac{2\tau}{4(\Delta+\Omega_{m})^{2}\tau^{2}+1}-\frac{2\tau}%
{4(\Delta-\Omega_{m})^{2}\tau^{2}+1}\right)  P
\end{multline}
This quantity can be recognized as the difference in intra-cavity energy of
anti-Stokes and Stokes modes, i.e. $\left\langle P_{m}\right\rangle
=(|a_{AS}|^{2}-|a_{S}|^{2})/T_{rt}$, as expected from energy conservation
considerations. Consequently, this analysis reveals how the mechanical mode
extracts (amplification) or loses energy (cooling) to the optical field,
despite the vastly different frequencies. Cooling or amplification of the
mechanical mode thus arises from the fact that the two side-bands created by
the harmonic mirror motion are \textit{not equal }in magnitude, due to the
detuned nature of the excitation (see Figure \ref{Figure2}). In the case of
cooling, a red detuned laser beam will cause the system to create more
anti-Stokes than Stokes photons entailing a net power flow from the mechanical
mode to the optical field. Conversely, a blue-detuned pump will reverse the
sideband asymmetry and cause a net power flow from the optical field to the
mechanical mode, leading to amplification. We note that in the case of
cooling, the mechanism is similar to the cooling of atoms or molecules inside
an optical resonator via "coherent scattering" as theoretically
proposed\cite{VuleticChu2000} and experimentally observed\cite{MaunzPuppe2004}%
. From the power calculation, the cooling or mechanical amplification rate
$\Gamma=\frac{\left\langle P_{m}\right\rangle }{m_{eff}\Omega_{m}%
^{2}\left\langle x^{2}\right\rangle }$ can be derived and expressed as:%
\begin{multline}
\Gamma=-\mathcal{F}^{2}\frac{8n^{2}\omega_{0}}{\Omega_{\mathrm{m}}m_{eff}%
c^{2}}C\times\label{E16}\\
\left(  \frac{1}{4(\Delta-\Omega_{m})^{2}\tau^{2}+1}-\frac{1}{4(\Delta
+\Omega_{m})^{2}\tau^{2}+1}\right)  P
\end{multline}
Here the finesse $\mathcal{F}$ has been introduced (as before) as well as the
previously introduced dimensionless coupling factor $C=\frac{\tau/\tau_{ex}%
}{4\Delta^{2}\tau^{2}+1}$which takes on a value between 0..1. Analysis of the
above formula allows derivation of the optimum detuning, for which the cooling
or amplification rate is maximum. Note that in the limit of $\kappa\gg
\Omega_{m}$ one recovers the result of the previous section (equation
\ref{Eq8}), as%
\begin{multline}
\underset{\Omega_{m}\rightarrow0}{\lim}\frac{1}{2\Omega_{m}}\left(  \frac
{1}{4(\Delta-\Omega_{m})^{2}\tau^{2}+1}-\frac{1}{4(\Delta+\Omega_{m})^{2}%
\tau^{2}+1}\right)  \label{E17}\\
=\frac{8\Delta^{2}\tau}{\left(  4\Delta^{2}\tau^{2}+1\right)  ^{2}}%
\end{multline}
Hence as noted before, in the weak retardation regime, the maximum cooling (or
amplification) rate for a given power occurs when the laser is detuned to the
maximum slope of the cavity lorenzian; i.e. $\left\vert \Delta\right\vert
=\kappa/2$, in close analogy to Doppler cooling\cite{Stenholm1986} in Atomic
Physics. On the other hand, in the resolved sideband regime optimum detuning
occurs when the laser is tuned either to the lower or upper \textquotedblleft
motional\textquotedblright\ sideband of the cavity, i.e. $\Delta=\pm\Omega
_{m}$. This behavior is also shown in Figure \ref{Figure7}. The above rate
modifies the dynamics of the mechanical oscillator. In the case of
amplification (blue detuned pump) it offsets the intrinsic loss of the
oscillator and overall mechanical damping is reduced. Ultimately, a
\textquotedblleft threshold condition\textquotedblright\ in which mechanical
loss is completely offset by gain occurs at a particular pump power. Beyond
this power level, regenerative mechanical oscillation occurs. This will be
studied in section 4 below.

For red detuning, the oscillator experiences enhanced damping. Beyond the
power flow analysis provided above, a simple classical analysis can also be
used to understand how such damping can result in cooling. Specifically, in
the absence of the laser, the mean energy (following from the equation for
$x$) obeys $\frac{d}{dt}\left\langle E_{m}\right\rangle =-\Gamma
_{m}\left\langle E_{m}\right\rangle +k_{B}T_{R}\Gamma_{m}$, implying that the
mean energy is given by the reservoir temperature $\left\langle E_{m}%
\right\rangle =k_{B}T_{R}$. When considering the modifications to this
equation resulting from back-action damping, an additional loss term appears
in the equation for the average energy:
\begin{equation}
\frac{d}{dt}\left\langle E_{m}\right\rangle =-(\Gamma_{m}+\Gamma)\left\langle
E_{m}\right\rangle +k_{B}T_{R}\Gamma_{m} \label{E18}%
\end{equation}

Note that within this classical model, the cooling introduced by the laser is
what has been described by some authors\cite{CohadonHeidmann1999} as
\textquotedblleft\textit{cold damping}\textquotedblright: the laser introduces
a damping without introducing a modified Langevin force (in contrast to the
case of intrinsic damping). This is a key feature and allows the enhanced
damping to reduce the mechanical oscillator temperature, yielding as a final
effective temperature for the mechanical mode under consideration:%
\begin{equation}
T_{eff}\cong\frac{\Gamma_{m}}{\Gamma_{m}+\Gamma}T_{R} \label{E19}%
\end{equation}
It deserves notice that the above formula predicts that arbitrarily low
temperatures can be attained. As discussed later, the laser damping also
introduces a small noise term, due to the quantum nature of light, which adds
a further Langevin force to previous equations. This will be considered in the
last section and is shown to provide the ultimate cooling limit of this
technique. Moreover, it is noted that the above formula is only
valid\cite{MarquardtChen2007, WilsonRaeNooshi2007} as long as $\Gamma\ll
\kappa$ \ and as $\frac{\Gamma_{m}}{\Gamma_{m}+\Gamma}>\frac{1}{Q_{m}}.$

For completeness, the in-phase component of the radiation pressure force is
also investigated. This component of the force causes a change in the
mechanical oscillator's rigidity, and its adiabatic contribution is the
well-known optical spring effect described earlier. Specifically, the change
in mechanical resonance frequency (from its intrinsic value) is given by:
\begin{multline}
\Delta\Omega_{m}=\mathcal{F}^{2}\frac{8n^{2}\omega_{0}}{\Omega_{\mathrm{m}%
}m_{eff}c^{2}}C\tau\times\label{Eq20}\\
\left(  \frac{\Delta-\Omega_{m}}{4(\Delta-\Omega_{m})^{2}\tau^{2}+1}%
+\frac{\Delta+\Omega_{m}}{4(\Delta+\Omega_{m})^{2}\tau^{2}+1}\right)  P
\end{multline}

Note that in the regime where the mechanical frequency is comparable to, or
exceeds the cavity decay rate it's behavior is quite different from that
described earlier for the conventional adiabatic case; and was only recently
observed experimentally\cite{SchliesserDelHaye2006}. As noted earlier, in the
adiabatic regime, the mechanical frequency is always downshifted by a
red-detuned laser (i.e. a reduced rigidity). However, when $\Omega
_{m}>1/2\cdot\kappa$ an interesting phenomena can occur. Specifically, when
the pump laser detuning is relatively small, the mechanical frequency shift is
opposite in sign to the conventional, adiabatic spring effect. The same
behavior can occur in the case of amplification ($\Delta>0$). Furthermore, a
pump detuning exists where the radiation-pressure induced mechanical frequency
shift is zero, even while the damping/amplification rate is non-zero. The
latter has an important meaning as it implies that the entire radiation
pressure force is \textit{viscous }for red detuning, contributing only to
cooling (or to amplification for blue detuning). Figure \ref{Figure7} shows
the attained cooling rate (as a contour plot) for fixed power and mechanical
oscillator frequency as a function of normalized optical cavity decay rate and
the normalized laser detuning (normalized with respect to $\Delta_{opt}%
=\sqrt{\Delta^{2}+\kappa^{2}/4}$. Evidently, the highest cooling/amplification
rates are achieved in this resolved sideband regime, provided the pump laser
is detuned to $+\Omega_{m}$ or $-\Omega_{m}$ (i.e. corresponding to the
cavities lower and upper motional sideband).

An important feature of cooling and amplification provided by dynamic
back-action is the high level of mechanical spectral selectivity that is
possible. Since, the cooling/amplification rates depend upon asymmetry in the
motional sidebands, the optical line-width and pump laser detuning can be used
to select a particular mechanical mode to receive the maximum cooling or
amplification. In effect, the damping/amplification rates shown above have a
spectral shape (and spectral maximum) that can be controlled in an
experimental setting as given by horizontal cuts of Figure \ref{Figure7}. This
feature is important since it can provide a method to control oscillation
frequency in cases of regenerative oscillation on the blue detuning of the
pump. Moreover, it restricts the cooling power to only one or a relatively
small number of mechanical modes. In cases of cooling, this means that the
overall mechanical structure can remain at room temperature while a target
mechanical mode is refrigerated.

\begin{figure}[ptbh]
\centering\includegraphics[width=7.6cm]{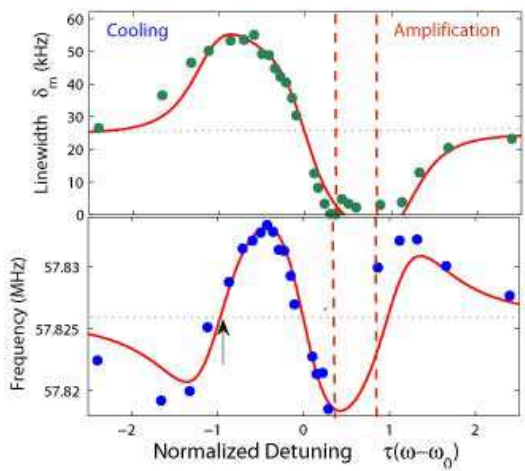}\caption{Dynamics in the
regime where $\Omega_{m}\sim\kappa$ as reported in
Ref.\cite{SchliesserDelHaye2006}. Upper panel shows the induced damping
/amplification rate ($\delta_{m}=\Gamma_{eff}/2\pi$) as a function of
normalized detuning of the laser at constant power. The points represent
actual experiments on toroidal microcavities, and the solid line denotes a fit
using the sideband theoretical model (Equations \ref{E16} and \ref{Eq20}).
Lower panel shows the mechanical frequency shift as a function of normalized
detuning. Arrow denotes the point where the radiation pressure force is
entirely viscous causing negligible in phase, but a maximum quadrature
component. The region between the dotted lines denotes the onset of the
parametric instability (as discussed in section 3). Graph stems from
Ref.\cite{SchliesserDelHaye2006}.}%
\label{Figure6}%
\end{figure}

The full-effect of the radiation-pressure force (both the in-phase component,
giving rise to a mechanical frequency shift, and the quadrature-phase
component which can give rise to mechanical amplification/cooling) were
studied in a \textquotedblleft detuning series\textquotedblright, as
introduced in the last section. The predictions made in the $\Omega
_{m}>0.5\kappa$ regime have been verified experimentally as shown in Figure
\ref{Figure6}. Specifically, the predicted change in the rigidity of the
oscillator was experimentally observed as shown in Figure \ref{Figure6} and is
in excellent agreement with the theoretical model (solid red line). Keeping
the same sample but using a different optical resonance with a line-width of
$113$ $%
\operatorname{MHz}%
$ ($57.8$ $%
\operatorname{MHz}%
$ mechanical resonance), the transition to a pure increasing and decreasing
mechanical frequency shift in the cooling and amplification regimes was
observed as shown in the preceding section, again confirming the validity of
our theoretical model based on the motional sidebands.

\section{Optomechanical Coupling and Displacement Measurements}
\subsection{Mechanical Modes of Optical Microcavities}
\begin{figure}[ptbh]
\centering\includegraphics[width=7.6cm]{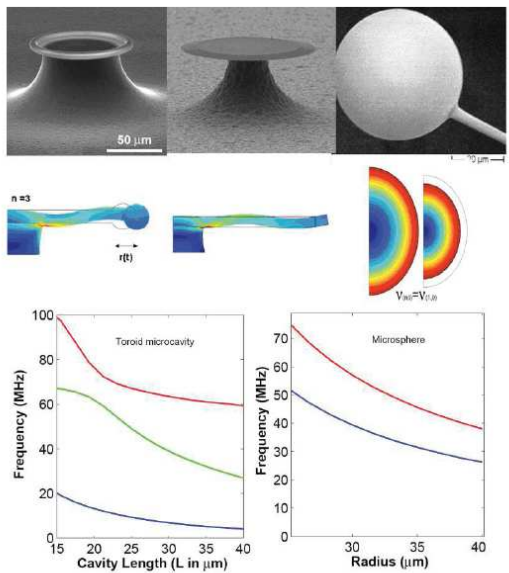}\caption{Upper panel: SEM
images and mechanical modes of several types of whispering gallery mode
microcavities: toroid microcavities\cite{ArmaniKippenberg2003}
microdisks\cite{KippenbergSpillane2003} and
microspheres\cite{BraginskyGorodetsky1989}. Also shown are the stress and
strain field in cross section of the fundamental radial breathing modes, which
include radial dilatation of the cavity boundary. Lower panel: the dispersion
diagram for the lowest lying, rotationally symmetric mechanical modes for a
toroid (as a function of its undercut) and for a microsphere (as a function of
radius).}%
\label{Figure8}%
\end{figure}

The coupling of mechanical and optical modes in a Fabry Perot cavity with a
suspended mirror (as applying to the case of gravitational wave detectors) or
mirror on a spring (as in the case of a cantilever) is easily seen to result
from momentum transfer upon mirror reflection and can be described by a
Hamiltonian formalism\cite{Law1995}. It is, however, important to realize that
radiation pressure coupling can also occur in other resonant geometries:
notably in the class of optical-whispering-gallery-mode (WGM) microcavities
such as microspheres\cite{BraginskyGorodetsky1989},
microdisks\cite{KippenbergSpillane2003} and
microtoroids\cite{ArmaniKippenberg2003} (see Figure \ref{Figure8}). Whereas in
a Fabry Perot cavity the momentum transfer occurs along the propagation
direction of the confined photons\cite{Braginsky1977}, the mechanism of
radiation pressure coupling in a whispering gallery mode cavity occurs normal
to the optical trajector\cite{CarmonRokhsari2005,
KippenbergRokhsari2005,RokhsariKippenberg2005}. This can be understood to
result from momentum conservation of the combined resonator-photon system as
photons, trapped within the whispering gallery by continuous total-internal
reflection, execute circular orbits. In particular, their orbital motion
necessitates a radial radiation pressure exerted onto the cavity boundary. The
latter can provide coupling to the resonator's mechanical modes.

While the mechanical modes of dielectric microspheres are well known and
described by analytic solutions (the Lamb theory), numerical finite-element
simulations are required to calculate the frequency, as well as strain and
stress fields, of more complicated structures (exhibiting lower symmetry) such
as micro-disks and micro-toroids. In the case of a
sphere\cite{MaSchliesser2007}, two classes of modes exist. Torsional
vibrations exhibit only shear stress without volume change, and therefore no
radial displacement takes place in these modes. Consequently these modes
cannot be excited using radiation pressure, which relies upon a change in the
optical path length (more formally, these modes do not satisfy the selection
rule of opto-mechanical coupling, which requires that the integral of
radiation pressure force and strain does not vanish along the optical
trajectory). In contrast, the class of modes for which volume change is
present is referred to as spheroidal modes. An example of this type of mode is
the radial breathing mode of a microsphere. In the case of a toroid, this
mode's equivalent is shown in Figure \ref{Figure8} and exhibits a
radially-symmetric, mechanical displacement of the torus. Note that in a
rotationally symmetric WGM microcavity, efficient opto-mechanical coupling
requires that the mechanical modes exhibit the same rotational symmetry (i.e.
satisfy the opto-mechanical selection rule). Deviations may occur in cases
where the symmetry is lowered further due to eccentricities in the shape of
the resonator, and in such cases excitation of modes of lowered symmetry is
possible. Excitation of eccentricity split modes has been observed in
microtoroids\cite{CarmonCross2007}, but these cases are not considered here
for the sake of simplicity.

A unique aspect of the micro-cavities, in contrast to other optomechanical
platforms, is that their fundamental mechanical breathing modes can exhibit
high mechanical frequency. The radial breathing mode of a $60-$micron-diameter
micro-sphere is equal to $\sim55$ $%
\operatorname{MHz}%
$, and, indeed, microwave-rate regenerative opto-mechanical oscillation has
been achieved in spheroidal microresonators\cite{CarmonVahala2007}. This
feature is important in the context of ground-state cooling as described
later. It is also potentially important should these devices find applications
as high-frequency oscillators. Another important aspect of the mechanical
modes in these structures is the level of dissipation, which is governed by
several loss mechanisms (clamping losses, thermoelastic losses, etc.). While
detailed understanding of dissipation in dielectric microcavities is presently
being established, quality factors as high as $50,000$ have been observed at
$50$ $%
\operatorname{MHz}%
$ and room temperature, comparing favorably with the best nano-mechanical
resonators\cite{NaikBuu2006} to date at low temperatures.

\begin{figure}[ptbh]
\centering\includegraphics[width=7.6cm]{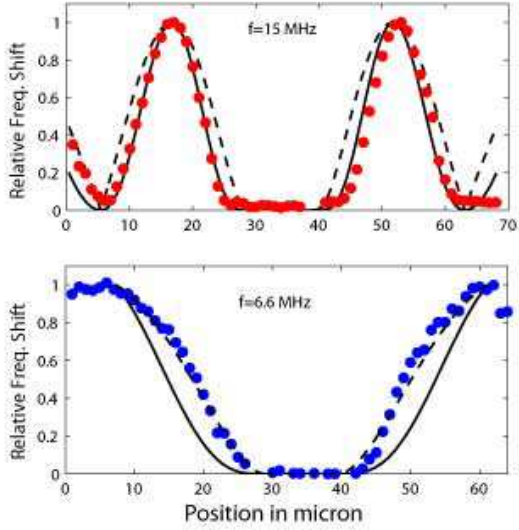}\caption{Scanning probe
microscopy of the two lowest lying micro-mechanical resonances of a toroid
microcavity. Lower graph: The normalized mechanical frequency shift for the
first mode as a function of position. Upper graph: The normalized frequency
shift for the second mechanical mode as a function of scanned distance across
the toroid. Superimposed is the scaled amplitude (solid line) and the
amplitude squared (dotted line) of the mechanical oscillator modes obtained by
finite element simulation of the exact geometry parameters (as inferred by
SEM). }%
\label{Figure9}%
\end{figure}

While the micro-mechanical modes coexist within the same physical structure as
the optical whispering gallery modes, it is important to note that there is
high level of spatial separation between the modes of these physical systems.
In fact, the optical whispering gallery confines optical energy to the extreme
periphery of the device, while the mechanical mode impacts the entire
structure. It is therefore possible to affect changes in the mechanical Q and
eigen-frequency spectrum by introduction of micro-mechanical probes without
affecting in any way the optical properties of the resonator. This feature can
provide additional ways to investigate the physics of these devices. For
example, a micro-mechanical probe, when scanned across the surface of a
microtoroid, is found to modify the eigenfrequency of a mechanical mode in
proportion to the amplitude-squared of the mode function. By measuring the
mechanical eigenfrequency during such a scan (using the optical probe
technique described above), the underlying mechanical mode can be
\textquotedblleft imaged.\textquotedblright\ Figure \ref{Figure9} provides
images taken of both a fundamental and first-excited mode in a microtoroid.

\subsection{Measuring the Opto-mechanical Response}
The mechanical properties of whispering gallery mode microcavities can be
probed by coupling resonant laser radiation into the microcavities using
tapered optical fibers\cite{SpillaneKippenberg2003}. Such probing will detect
mirror or cavity motion as a modulation in the power transmitted past the
resonator. This modulation can, in turn, be measured as a photocurrent upon
detection with a photodiode. The continuous, optical probe wave, itself, can
also be used to affect changes in the mirror dynamics via the back-action
effect. A schematic of the experiment, which can be used to study
optomechanical phenomena, is shown in the Figure \ref{Figure10} below. It
consists of a continuous-wave pump laser (here, a 965$-$nm diode laser) which
is coupled into a standard, single-mode optical fiber. This fiber enters the
experimental chamber, where it also contains a tapered region used to enable
evanescent coupling between the tapered fiber and various types of
microcavities. The output fiber is connected to different analysis
instruments, including an electrical spectrum analyzer and oscilloscope. Note
that moreover locking electronics is used to ensure operation at a fixed
detuning.\begin{figure}[ptbh]
\centering\includegraphics[width=7.6cm]{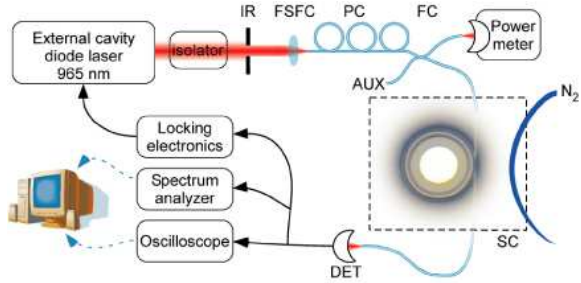}\caption{Experimental
setup for the observation of cavity cooling or oscillation of a mechanical
oscillator. All relevant data from the electronic spectrum analyzer and the
oscilloscope are transferred to a computer controlling the experiment. More
details in the text. IR: iris, FSFC: free-space to fibre coupler, PC: fibre
polarization controller, FC: fibre coupler, AUX: auxiliary input, SC: sealed
chamber, DET: fast photoreceiver.}%
\label{Figure10}%
\end{figure}

\subsection{Displacement Sensitivity}
Both the damping rate as well the effective temperature of a mechanical mode
are observed by measuring the calibrated displacement noise spectra as a
function of power. Despite their small amplitude, these thermally excited
oscillations are readily observable in the transmitted light. Indeed, optical
interferometers are among the most sensitive monitors for displacement. For
quantum-noise-limited homodyne detection the shot noise limited displacement
sensitivity of a cavity opto-mechanical system is given by:%
\begin{equation}
\delta x_{\min}\cong\frac{\lambda}{8\pi\mathcal{F}\sqrt{\eta P/\hslash\omega}}%
\end{equation}

\begin{figure}[ptbh]
\centering\includegraphics[width=7.6cm]{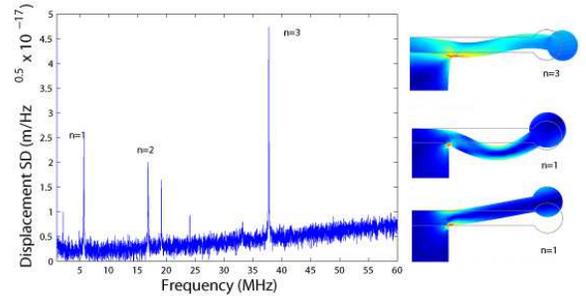}\caption{Calibrated\cite{SchliesserRiviere2007}
displacement spectral density as measured by the setup shown in Figure 10. The
peaks denote different mechanical eigenmodes of the toroidal microcavity. The
probe power is sufficiently weak such that the mechanical modes amplitude is
dominated by Brownian motion at room temperature and backaction effects are
negligible. Cross-sectional representations of the $n=1,2,3$ modes and their
corresponding spectral peaks are also given as inferred by finite element
simulations.}%
\label{Figure11}%
\end{figure}

For numbers typical of the toroidal microcavity work ($\mathcal{F\approx
}40000$, $\eta\approx0.5,P$ $\approx1$ $\mu%
\operatorname{W}%
,$ $\lambda=1064$ $%
\operatorname{nm}%
$) this implies a displacement sensitivity of $\delta x_{\min}\cong
5\cdot10^{-19}%
\operatorname{m}%
/\sqrt{%
\operatorname{Hz}%
}$. Note that this is a remarkably low level, which has been experimentally
achieved in a similar approach at the LKB\cite{ArcizetCohadon2006}.
Interestingly, it is, and, in principle, even sufficient to detect the zero
point motion $\delta x_{\min}\cong5\cdot10^{-16}%
\operatorname{m}%
$ within a 1 kHz resolution bandwidth. In practice, however, such a value can
only be achieved in cases where true quantum-limited-readout is present,
necessitating lasers which operate with quantum limited amplitude and phase
noise, i.e., do not have excess classical noise in either of the two
quadratures (the latter is the case of Nd:YAG Lasers for frequencies above ca.
$1$ $%
\operatorname{MHz}%
$). For the experimental measurements described\cite{SchliesserDelHaye2006}
herein the actual displacement sensitivity did not achieve this level owing to
the fact that the diode lasers exhibited excess phase noise, and limited the
sensitivity to a value of about $\delta x_{\min}\cong5\cdot10^{-18}%
\operatorname{m}%
/\sqrt{%
\operatorname{Hz}%
}$. Recent work, however, has also obtained higher sensitivity by employing
low noise lasers\cite{SchliesserRiviere2007}. A typical calibrated and
broadband displacement spectrum that can be attained with diode lasers is
shown in Figure \ref{Figure11}. It reveals several mechanical modes, which can
be accurately assigned via finite element modeling. 

\section{Blue Detuning: Mechanical Gain and Parametric Oscillation
Instability}
\subsection{Threshold and Mode Selection Mechanisms}
Regenerative oscillation occurs in the case of a blue-detuned optical pump
wave and at pump power levels above a threshold pump power $(P_{thresh}).$ The
oscillation threshold occurs when mechanical gain balances mechanical loss. By
equating the expression for the mechanical amplification rate $\Gamma$ of a
specific mechanical mode (equation \ref{E16}) to its intrinsic loss
($\Gamma_{m}=\Omega_{m}/Q_{m}$), the following expression for threshold
optical pump power results.%
\begin{multline}
P_{thresh}=\frac{\Omega_{m}^{2}}{Q_{m}}\frac{m_{eff}c^{2}}{\omega
_{0}\mathcal{F}^{2}8n^{2}C}\times\label{EqParamThresh}\\
\left(  \frac{1}{4(\Delta-\Omega_{m})^{2}\tau^{2}+1}-\frac{1}{4(\Delta
+\Omega_{m})^{2}\tau^{2}+1}\right)  ^{-1}%
\end{multline}

In the case of weak retardation ($\kappa\gg\Omega_{m}$) this result simplifies
to:%
\begin{equation}
P_{thresh}\underset{\kappa\gg\Omega_{m}}{=}\frac{\Omega_{m}^{3}}{Q_{m}}%
\frac{m_{eff}c^{3}}{\omega_{0}\mathcal{F}^{3}8n^{3}RC}\left(  \frac
{8\Delta\tau}{4\Delta^{2}\tau^{2}+1}\right)  ^{-1} \label{EqparamThresh2}%
\end{equation}

It is worth noting that even though cooling does not exhibit a similar
threshold condition, the above condition, when expressed for the case of
red-detuning of the pump wave, gives the condition in which the radiation
pressure cooling rate equals the heating rate of the mechanical mode. As such,
a factor of \ $\times2$ in temperature reduction (cooling) already requires
that parametric oscillation is observable for the corresponding blue detuning.

\begin{figure}[ptbh]
\centering\includegraphics[width=7.6cm]{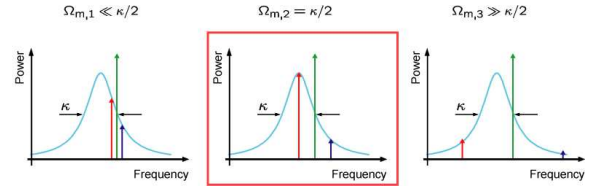}\caption{Back-action
tuning for mode selection for a fixed laser detuning corresponding to
$\Delta=\kappa/2$. The target mode that receives maximum gain (or optimal
cooling for\ $\Delta=-\kappa/2$) can be controlled by setting the cavity
linewidth to produce maximum sideband asymmetry for that particular mechanical
mode. In this schematic, three mechanical modes (having frequencies
$\Omega_{\mathrm{m,i}}$, $i=1,2,3)$ interact with an optical pump interact
with the optical cavity mode, however, in the present scenario only the
intermediate mode experiences maximum gain (or cooling) since its sideband
asymmetry is maximal (since $\Omega_{\mathrm{m,2}}/\kappa=0.5$). It is
important to note however, that if the laser detuning is allowed to vary as
well, the highest frequency mode would experience the the largest gain if
$\Delta=\Omega_{\mathrm{m,2}}$ was chosen.}%
\label{Figure12}%
\end{figure}

An important feature of the mechanical gain is its dependence with respect to
the mechanical eigen-frequency. Mechanical modes whose eigenfrequencies fall
near the peak of this curve have the lowest threshold pump power for
oscillation. The general shape of this curve can be inferred from
\ref{Figure7}, which shows a contour plot of both the gain and the cooling
rate versus the normalized mechanical eigenfrequency and the normalized
detuning. As noted earlier above, horizontal slices of this plot give the gain
(or cooling rate) spectral shapes. Experimental control of the spectral peak
of the gain (cooling rate) is obviously important since it determines which
mechanical modes oscillate or receive maximum cooling. The contour of maximum
gain appears as a dashed contour in the plot (likewise there is a
corresponding contour for maximum cooling rate). The unresolved sideband case
(case of weak retardation) in the figure provides an convenient physical limit
in which to illustrate one form of spectral control. In this case, as noted
before, the contour of maximum gain (cooling) occurs when the pump wave is
detuned to the half-max position of the optical lineshape function (cf. the
vertical axis value of the contour in unresolved sideband regime). The maximum
mechanical gain increases along this contour as the parameter $\Omega
_{\mathrm{m}}/\kappa$ increases towards $0.5$ and then diminishes beyond this
value (for pump-wave detuning fixed). (Beyond this value, even somewhat before
it is reached, the pump detuning must be adjusted continuously in concert with
increases in $\Omega_{\mathrm{m}}/\kappa$ to remain on the contour of maximum
gain. This is a result of transition into the sideband resolved regime as
described earlier). This behavior can be understood in the context of the
motional side band (Stokes/anti-Stokes waves) description provided earlier
(see Figure \ref{Figure2}). Specifically, the case of a pump wave detuned to
the half-max point is diagrammed in figure \ref{Figure12} for three values of
the parameter $\Omega_{\mathrm{m}}/\kappa$ ($<,=,>0.5$). The corresponding
side-band configuration in each of these cases is also illustrated for
comparison. Since mechanical gain is largest when the sideband asymmetry is
maximum, the intermediate case of $\ \Omega_{\mathrm{m}}/\kappa=0.5$ will
exhibit maximum gain in the scenario depicted in Figure \ref{Figure12}.
Through adjustments of the optical mode line-width (as can be done by
controlling waveguide loading of the microresonator) an optimum $\Omega
_{\mathrm{m}}/\kappa$ can be set experimentally. This method has, in fact,
been used to provide targeted oscillation of mechanical modes (even into the
microwave regime) through control of optical waveguide loading. It is
important to note however that the above considerations assume a constant
normalized detuning. If the detuning is allowed to vary as well, maximum
cooling or amplification rates always occurs in the resolved sideband regime,
when the detuning equates to $\Delta=\pm\Omega_{\mathrm{m}}$(cf. Figure
\ref{Figure7} and \ref{Figure12} ). The inherent advantages of this regime are
that it enables a higher level of asymmetry in the side bands. As described in
the next section, side-band asymmetry takes on even greater significance in
the context of ground state cooling.

\subsection{Threshold dependence on optical Q and mechanical Q}
\begin{figure}[ptbh]
\centering\includegraphics[width=7.6cm]{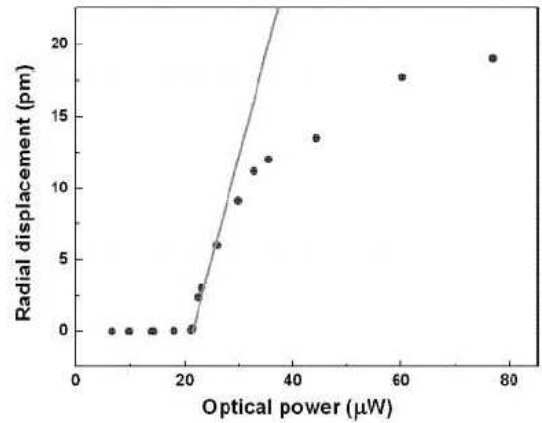}\caption{Regenerative
oscillation amplitude plotted versus pump power. The threshold knee is clearly
visible. In this case a threshold of $20$ $\mu\operatorname{W}$ is observed.
Figure from reference\cite{KippenbergRokhsari2005}.}%
\label{Figure13}%
\end{figure}

There are several additional features of the threshold equation that
are worth noting. First, it exhibits a classic inverse dependence on
mechanical Q. This is a signature for any regenerative system.
Second, in the unresolved sideband regime, the threshold exhibits an
inverse-cubic dependence on the optical Q factor (and
correspondingly also Finesse). The measurement of these inherent
dependences provides further evidence of the underlying nature of
the interaction. Measurement of the threshold is straightforward and
involves monitoring the photocurrent of the detected transmission
either in the time domain or on a spectrum analyzer (as described
earlier). The amplitude of oscillations at a particular mechanical
eigen-frequency will exhibit a \textquotedblleft threshold
knee\textquotedblright\ when plotted versus the coupled optical pump
power. This knee is an easily measurable feature and one example is
provided in Figure \ref{Figure13}. The first observation of
radiation pressure parametric oscillation instability, as predicted
by Braginsky, was made in toroidal
micro-cavities\cite{CarmonRokhsari2005,KippenbergRokhsari2005,
RokhsariKippenberg2005}. The setup employed a tapered optical fiber
coupling such that optical loading could be controlled during the
experiment. This enabled control of the optical linewidth as
described above to effect control of the specific mechanical mode
designated for oscillation. When exciting the cavity using a
blue-detuned laser pump, an oscillatory output of the cavity could
be observed, indicative of the excitation of mechanical modes. The
oscillation is readily observable in microtoroids with threshold
values in the microwatt range. Indeed, by using typical parameters
in the above threshold
equation ($\mathcal{F\approx}10^{5},$ $m_{eff}\approx10^{-11}%
\operatorname{kg}%
,$ $Q_{m}\approx10^{4}$) a threshold in the range of a few micro-Watts is
predicted (which is even below the threshold for
Raman\cite{SpillaneKippenberg2002} and parametric
oscillations\cite{KippenbergSpillane2004}). The fact that this process can be
observed in a reproducible manner allows probing of fundamental metrics of
these phenomena including the above-noted mechanical and optical Q-dependences.

\begin{figure}[ptbh]
\centering\includegraphics[width=7.6cm]{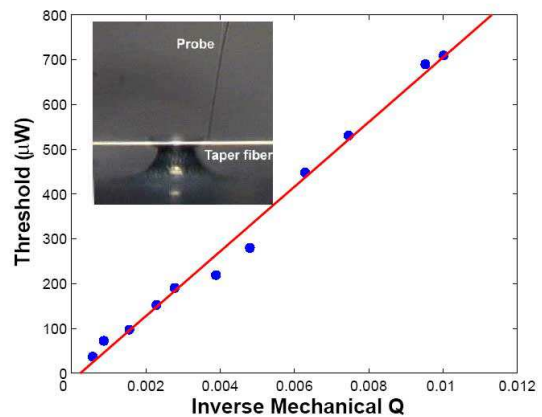}\caption{Main figure: The
observed threshold for the parametric oscillation (of an $n=1$ mode) as a
function of inverse mechanical quality factor. In the experiment, variation of
$Q$ factor was achieved by placing a fiber tip in mechanical contact with the
silica membrane, which thereby allowed reduction of the mechanical $Q$ (cf.
inset). The mechanical mode was a $6$ $\operatorname{MHz}$ flexural mode. }%
\label{Figure14}%
\end{figure}

Measurement of the dependence of the parametric oscillation threshold on the
mechanical $Q$ is shown in Figure \ref{Figure14}. In this experiment, first
reported in ref\cite{KippenbergRokhsari2005}, the mechanical $Q-$factor was
varied while the optical Q factor was left unchanged. In a variation on the
mechanical probing technique described above to provide image scans of the
mechanical mode, a micro-probe, in the form of a sharp silica fiber tip, was
brought into contact, at a fixed position, with the interior of the
microcavity. This caused dissipative coupling of the mechanical mode,
decreasing its value from an initial, room temperature and ambient pressure
$Q-$value of 5000 to below 50. As noted above, while the mechanical probe
modifies the $Q$ (and weakly, the eigen-frequency) of the mechanical mode, it
has no effect on the optical performance of the whispering gallery. Therefore,
by using this method, the dependence of threshold on mechanical Q can be
probed in a nearly ideal way. Gradual change could be induced by variation of
the tip pressure. Figure \ref{Figure14} shows the result of this measurement
for the $n=1$ flexural mode. As is evident, there is excellent agreement with
the theoretical prediction.

\begin{figure}[ptbh]
\centering\includegraphics[width=7.6cm]{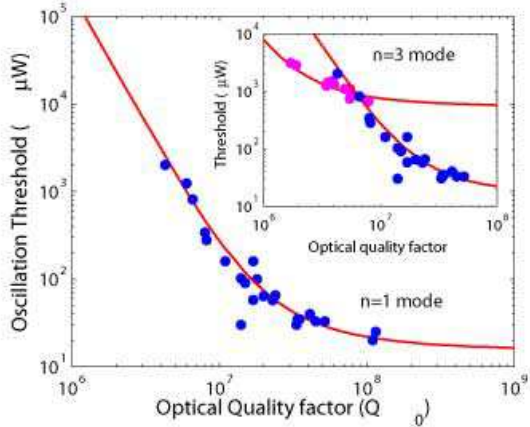}\caption{Main panel shows
the measured mechanical oscillation threshold (in micro-Watts) from
Ref.\cite{KippenbergRokhsari2005} plotted versus the optical Q factor for the
fundamental flexural mode ($n=1,$ $\Omega_{m}/2\pi=4.4\operatorname{MHz},$
$m_{eff}\approx3.3\times10^{-8}\operatorname{kg},$ $Q_{m}\approx3500$). The
solid line is a one-parameter theoretical fit obtained from the minimum
threshold equation by first performing a minimization with respect to coupling
($C$) and pump wavelength detuning ($\Delta$), and then fitting by adjustment
of the effective mass. Inset: The measured threshold for the 3$^{rd}$ order
mode ($n=3,$ $\Omega_{m}/2\pi=49\operatorname{MHz},$ $m_{eff}\approx
5\times10^{-11}\operatorname{kg},$ $Q_{m}\approx2800$) plotted versus optical
Q. The solid line gives again the theoretical prediction. The $n=1$ data from
the main panel is superimposed for comparison. Figure stems from
Ref.\cite{KippenbergRokhsari2005}.}%
\label{Figure15}%
\end{figure}

Next, we examine the measured oscillation threshold dependence on the optical
Q factor. To illustrate behavior occurring in the sideband resolved and
unresolved regimes two mechanical modes were measured: a fundamental $(n=1)$
at $4.4%
\operatorname{MHz}%
$ and a third-order mode $(n=3)$ at $49$ $%
\operatorname{MHz}%
$. The optical Q factor was adjusted by exciting different radial and
transverse optical modes. For lower optical $Q$, wherein the mechanical
oscillation frequency falls within the cavity bandwidth (i.e. the adiabatic
regime), a rapid dependence $1/Q^{3}$ is observed, which is in agreement with
the threshold equation (i.e. $P_{thresh}\propto1/\mathcal{F}^{3}$ cf. equation
\ref{EqparamThresh2}). However, the scaling of threshold changes once a
transition from the unresolved (weak retardation) to the resolved sideband
limit occurs. Indeed, for $\Omega_{m}\gtrsim\kappa$ , the threshold dependence
on optical $Q$ (and Finesse) weakens and eventually approaches an asymptotic
value for $\Omega_{m}\gg\kappa$. The deviation from the cubic dependence is
indeed observed experimentally as shown in Figure \ref{Figure15}. The solid
line in the figure is a prediction based on the threshold equation
\ref{EqParamThresh} with effective mass as an adjustable parameter (where
optimum detuning is assumed, and optimum coupling for each optical $Q$ value).
The inset of Figure \ref{Figure15} shows the threshold behavior for the $n=3$
mechanical mode which for which $\Omega_{m}\gtrsim\kappa$ is satisfied for the
entire range of $Q$ values. As expected, a much weaker dependence on optical
$Q$-factor is found as predicted theoretically. Direct comparison with the
$n=1$ mode data shows that oscillation on the $n=3$ mode is preferred for
lower optical Qs. Indeed, preference to the $n=3$ mode was possible by
increased waveguide loading of the microcavity in agreement with theory. The
solid curve in the inset gives the single-parameter fit to the $n=3$ data
yielding , $m_{eff}=5\times10^{-11}%
\operatorname{kg}%
$ which is a factor of 660 lower than the mass of the $n=1$ mode.

\subsection{Oscillation Linewidth}
A further important characteristic of the regenerative mechanical oscillation
is the linewidth of the mechanical oscillation frequency. Theoretically, the
limit for the line-width in the case of temperatures $k_{B}T_{R}>\hslash
\Omega_{m}$is set by classical, thermal noise and obeys the
relationship\cite{HosseinZadehRokhsari2006, HosseinZadehVahala2007}:
\[
\Delta\Omega_{m}\cong\frac{k_{B}T_{R}}{P}\left(  \frac{\Omega_{m}}{Q_{m}%
}\right)  ^{2}%
\]

\begin{figure}[ptbh]
\centering\includegraphics[width=7.6cm]{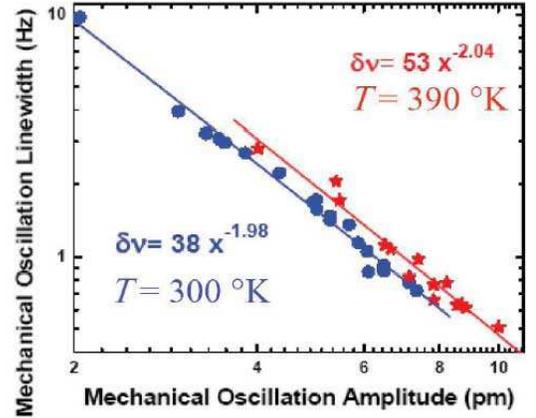}\caption{Line-width
measurements from Refs.\cite{HosseinZadehRokhsari2006,HosseinZadehVahala2007}
of the opto-mechanical oscillator for different amplitudes of oscillation
plotted in picometers. The measurement is done at room temperature (dots) and
at temperature 90 ${{}^{\circ}}$C above room temperature (stars). The solid
lines and the corresponding equations are the best fits to the log-log data.
Solid line denotes theoretically expected behavior. }%
\label{Figure16}%
\end{figure}

Here P is the power dissipated in the mechanical oscillator, i.e. $P=$
$\Gamma_{m}m_{eff}\Omega_{m}^{2}\left\langle x^{2}\right\rangle $.
Consequently the equation predicts that the line-width and mechanical
amplitude satisfy $\Delta\Omega_{m}\propto1/\left\langle x^{2}\right\rangle $.
Indeed, experimental work has confirmed this scaling in toroidal
microcavities\cite{HosseinZadehRokhsari2006,RokhsariHosseinZadeh2006}.The
measured, inverse-quadratic dependence (as first reported in
Refs.\cite{HosseinZadehRokhsari2006,HosseinZadehVahala2007} is presented in
Figure \ref{Figure16}. Fundamental line-widths that are sub Hertz have been
measured, however, with improvements in mechanical Q factor, these values are
expected to improve, owing to the inverse-quadratic dependence appearing in
the above formula.

\section{Red Detuning: Radiation Pressure Cooling}
\subsection{Experimental Setup}
From the theoretical section it is clear that both cooling and amplification
are indeed the same effect, albeit with opposite signs in the work performed
on the mechanical system. Cooling was first theoretically noted by
Braginksy\cite{BraginskyVyatchanin2002}, who proposed that red detuning could
be used to \textquotedblleft tranquilize\textquotedblright\ the mechanical
modes of an interferometer. It has subsequently been considered in other
work\cite{MetzgerKarrai2004,KippenbergRokhsari2005}. While the manifestations
of parametric amplification are readily observed (leading to a strong periodic
modulation of the cavity transmission), the effect of cooling is more subtle
as it decreases the Brownian motion of the mechanical motion. Consequently,
the observation of cooling requires both a careful calibration of the
mechanical displacement spectra and sufficient signal-to-noise to detect
the\textit{ cooled} mechanical mode. Indeed, cooling was observed only after
the parametric oscillation in a series of three experiments reported at the
end of 2006, which employed coated micro-mirrors (in the Paris
\cite{ArcizetCohadon2006Nature} and the Vienna\cite{GiganBohm2006}
experiments) and toroidal micro-cavities\cite{SchliesserDelHaye2006} (at the
MPQ in Garching). As mentioned earlier, there is no threshold condition for
cooling (as in the case of the parametric oscillation), however, to achieve
efficient temperature reduction the power injected when red detuning, must
greatly exceed the threshold of the parametric oscillation instability for
blue detuning.

Cooling in toroidal microcavities was observed by pumping with an external
diode laser that is red detuned with respect to the optical resonance. It
deserves notice that this detuning is intrinsically unstable owing to the
thermal bi-stability effect\cite{CarmonYang2004} . The cavity absorption
induced heating causes a change of the cavity path length, which subsequently
causes a red-shift of the cavity resonance. This feature leads to thermal,
self-locking on the blue-sideband used to observe parametric amplification and
oscillation, but simultaneously destabilizes locking on the red sideband. This
has necessitated the implementation of fast electronic feedback to the laser,
in order to be able to observe cooling. As an error signal, either the
detected transmission signal was used directly or a signal derived from this
by a frequency modulation technique\cite{BjorklundLevenson1983}. The error
signal is pre-amplified with a low-noise amplifier (DC - 1 $%
\operatorname{MHz}%
$), the two outputs of which are fed to two, custom-built
proportional-integral controllers with bandwidths on the order of 1 $%
\operatorname{kHz}%
$ and 1 $%
\operatorname{MHz}%
$. Both controllers allow us to apply an offset to the error signal input,
enabling continuous variation of the control set-point and thus constant
detuning from line center. Without further amplification, the output of the
slower controller is applied to a piezoelectric element actuating the grating
in the laser to tune the laser emission frequency. For the compensation of
fast fluctuations, the output from the faster controller is applied to a field
effect transistor parallel to the laser diode. The consequent temporary change
of diode current leads to the desired laser frequency adjustment. Laser
emission power is affected only on the order of 5\% and, since the output of
the fast controller is high-pass filtered (cut-off $\gg$ 10 Hz), it remains
unmodified on average.

\begin{figure}[ptbh]
\centering\includegraphics[width=7.6cm]{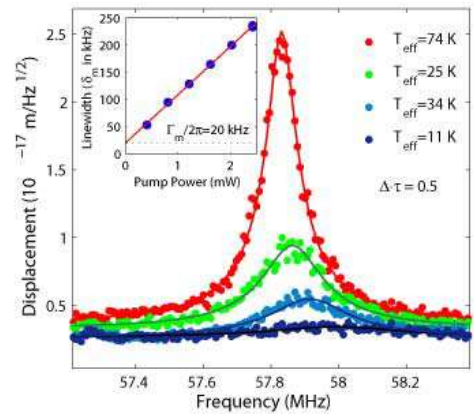}\caption{Main figure
shows the normalized, measured noise spectra around the mechanical breathing
mode frequency for $\Delta\cdot\tau\approx-0.5$ and varying power
($0.25,0.75,1.25,$and $1.75$ $\operatorname{mW}$). The effective temperatures
were inferred using mechanical damping, with the lowest attained temperature
being $11$K. (b) Inset shows increase in the linewidth (damping $\delta
_{m}=\Gamma_{eff}/2\pi$) of the $57.8-\operatorname{MHz}$ mode as a function
of launched power, exhibiting the expected linear behavior as theoretically
predicted. From reference\cite{SchliesserDelHaye2006}.}%
\label{Figure17}%
\end{figure}

The described locking technique was used in Ref.\cite{SchliesserDelHaye2006}
to observe back-action cooling by radiation pressure. In the experiment the
laser was detuned to $\Delta\cdot\tau\approx-0.5$. The mechanical mode under
consideration in this study is a radial breathing mode with resonance
frequency of 57.8 $%
\operatorname{MHz}%
$. Since the mechanical mode is in thermal equilibrium with the environment at
room temperature (T =300 K), it follows from the classical equipartition
theorem that the mechanical oscillator undergoes Brownian motion.

\subsection{Experimental Observation of Cooling}
Given the high displacement sensitivity, it is possible to record
important characteristics of the mechanical mirror motion, such as
the resonance frequency $\Omega_{m}/2\pi$, damping rate and
displacement spectral density $S_{x}(\omega)$. The determination of
the intrinsic mechanical properties was carried out by injecting
very-low power into the cavity, and recording the power spectral
density with an electrical spectrum analyzer. Upon sufficiently low
injected power, this leads to a nearly identical noise spectra for
both blue and red detuning for $\Delta=\pm\kappa/2$ (since the
mechanical amplification or cooling rate is much smaller than the
intrinsic mechanical damping rate). Conversion to displacement
spectral density then requires knowledge of the effective
mass\cite{PinardHadjar1999} of the mechanical mode which could be
independently determined in two ways. First, from finite-element
simulation of the actual cavity geometry parameters, as inferred
from scanning electron microscopy images. As detailed in
Ref.\cite{KippenbergRokhsari2005} this is accomplished via the
relation $m_{eff}=\frac{E_{m}}{\Omega_{m}^{2}(\delta R)^{2}}$, where
$\delta R$ is the mechanical energy causing a displacement in the
radial direction. Second, the effective mass was determined by
experimental measurements, by recording the threshold for parametric
oscillation for blue detuning and subsequently inverting the
threshold equation for the effective mass. Note that in the
described experiments, both techniques agree very well, from which
the effective mass of the radial breathing mode of Figure
\ref{Figure17} is
inferred to be $m_{eff}=1.5\times10^{-11}%
\operatorname{kg}%
$ . Correspondingly, the rms motion caused by Brownian motion of the radial
breathing mode at room temperature is on the order of $\left\langle
x^{2}\right\rangle ^{1/2}\approx5\times10^{-14}%
\operatorname{m}%
$.

Figure 18 shows the displacement spectral density for the radial breathing
mode under conditions of constant red detuning ($\Delta=-\kappa/2$) for
varying input power levels $P$. By extrapolating the resonant frequency and
linewidth to zero power (cf. Inset) the intrinsic resonance frequency and an
intrinsic mechanical Q factor of $2890$ were extracted. Note that much higher
mechanical $Q$ $(>50,000)$ are possible in an evacuated
chamber\cite{MaSchliesser2007} and by optimizing the micro-cavity shape to
reduce dissipative clamping losses. In the reported experiment the optical
line-width was $50$ $%
\operatorname{MHz}%
$, equivalent to an optical quality factor of $4.4\times10^{6}$. When varying
the pump power, a clear reduction of the noise spectra is observed as shown in
the main panel. The shape of the spectra changes in two dramatic ways. First
the line-width of the mechanical spectra increase, owing to the fact that the
light field provides a viscous force, thereby increasing the damping rate.
Note that this rate varies linearly with applied power (cf. inset). Moreover,
and importantly, the areas of the mechanical curves also reduce, which are a
direct measure of the mechanical breathing modes (RBM) temperature. Indeed the
effective temperature of the RBM is given by $k_{B}T_{eff}=\int\Omega
^{2}m_{eff}S_{x}(\Omega)d\Omega$ and correspondingly the peak of the
displacement spectral density is reduced in a quadratic fashion with applied
laser power. While measurement of the calibrated noise spectra is the most
accurate way to determine the effective temperature, the simplified analysis
presented earlier (and neglecting any other heating mechanism) yields a
temperature (cf. Equation \ref{E19}) given by the ratio of damping rates (with
and without the pump laser):
\begin{equation}
T_{eff}\cong\frac{\Gamma_{m}}{\Gamma_{m}+\Gamma}T_{R}%
\end{equation}

As noted before this formula only proves valid when $\frac{\Gamma_{m}}%
{\Gamma_{m}+\Gamma}<1/Q_{m}$ and for cooling rates satisfying $\Gamma<<\kappa
$. Note that the maximum temperature reduction factor is bound by the cavity
decay rate $\symbol{126}\Gamma/\kappa$. Thus, the highest temperature
reduction factor which can be attained is given by $\sim\kappa/\Gamma_{m}$.
Furthermore, additional modifications of this expression are necessary when
entering the limit of temperatures, which correspond to only a few quanta, and
will be considered in the next section.

For the highest pump power ($2%
\operatorname{mW}%
$ and $970$ $%
\operatorname{nm}%
$), the effective temperature was reduced from 300 K to 11 K. This experiment,
reported in Ref.\cite{SchliesserDelHaye2006}, along with two earlier
publications\cite{ArcizetCohadon2006Nature,GiganBohm2006} represent the first
demonstration of radiation-pressure back-action cooling. It is noted for
completeness that the experiment described in Ref. \cite{GiganBohm2006} also
attributed an appreciable cooling effect due to thermal effects in the mirror
coatings. Indeed, of the physical mechanisms that can create opto-mechanical
coupling temperature is another possibility. Temperature variations introduced
by absorption, for example, create a well-known, trivial coupling, by way of
thermal expansion (and absorption of photons). Indeed both mechanical
amplification\cite{ZalalutdinovZehnder2001} and
cooling\cite{MetzgerKarrai2004} have been demonstrated using this mechanism.
In the present, microtoroid studies, thermal effects are negligible. This is
known, first, because of the parametric-instability studies of the previous
section, where both the observed threshold dependence on optical Q as well as
the magnitude of the threshold power, itself, are in excellent agreement with
the theory of radiation-pressure-induced coupling. Second, the recent
observation of microwave-rate parametric oscillations\cite{CarmonVahala2007}
confirms the broad-band nature of the underlying mechanism, greatly exceeding
bandwidths possible by thermal coupling mechanisms.

\begin{figure}[ptbh]
\centering\includegraphics[width=7.6cm]{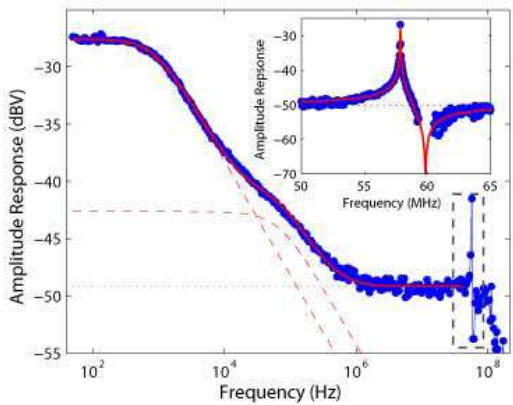}\caption{The frequency
response from $0-200\operatorname{MHz}$ of a toroidal opto-mechanical system,
adopted from Ref.\cite{SchliesserDelHaye2006}. The plateau occurring above
$1\operatorname{MHz}$ is ascribed to the (instantaneous) Kerr nonlinearity of
silica (dotted line). The high-frequency cutoff is due to both detector and
cavity bandwidth. The response poles at low frequency are thermal in nature.
Inset: Data in the vicinity of mechanical oscillator response which shows the
interference of the Kerr nonlinearity and the radiation pressure-driven
micromechanical resonator (which, on resonance, is $\pi/2$ out-of phase with
the modulating pump and the instantaneous Kerr nonlinearity). From the fits
(solid lines) it can be inferred that the radiation pressure response a factor
of $260$ $\,$larger than the Kerr response and a factor of $\times100$ larger
than the thermo-mechanical contribution.}%
\label{Figure18}%
\end{figure}

To \textit{quantitatively} assess the contributions of thermal and
radiation-pressure induced effects in the present context, pump-probe-type
response measurements were performed using a second laser coupled to the
cavity and operating in the wavelength region around $1550$ $%
\operatorname{nm}%
$. This laser serves as \textit{pump}, providing a sinusoidally modulated
input power $P(t)=P_{0}(1+\epsilon\sin(\Omega t))$, which in turn causes the
optical resonances to periodically shift via both thermal effects and
radiation-pressure-induced mechanical displacement, but also via the
Kerr-nonlinearity of fused silica. These shifts are then \textquotedblleft
read out\textquotedblright\ with the $965$ $%
\operatorname{nm}%
$ probe laser tuned to the wing of an optical resonance. Modulating the power
of the pump laser and demodulating the detected probe power with the same
frequency, it is possible to measure the micro-cavity response caused by all
three nonlinearities (thermal, Kerr, and radiation pressure). Note that due to
their different spectral response, the three nonlinearities can be readily
differentiated when performing a frequency sweep. While electronic modulation
and demodulation are conveniently accomplished with an electronic network
analyzer, we use a fibre-coupled interferometric LiNbO$_{3}$ amplitude
modulator to generate the modulated pump light. Data were taken on a
logarithmic frequency scale between $50%
\operatorname{Hz}%
$ and $200$ $%
\operatorname{MHz}%
$, and, subsequently, on a linear frequency scale in the interval $50-65$ $%
\operatorname{MHz}%
$. From finite-element simulation, this resonance can be identified as the
radial breathing mode. The result of this measurement, as first reported in
Ref.\cite{SchliesserDelHaye2006}, is shown in figure \ref{Figure18}. Several
features are evident from the graph and are now discussed in detail.

First, the plateau in the response, for frequencies beyond $1$ $%
\operatorname{MHz}%
$ and up to the cavity (and detector) cut-off at $200%
\operatorname{MHz}%
$, is attributed to the intensity-dependent refractive index (Kerr effect) of
fused silica\cite{TreussartIlchenko1998,RokhsariVahala2005}. Importantly,
since both thermal and mechanical responses exceed the Kerr non-linearity in
some frequency domains, the Kerr effect (plateau) provides a precisely known
reference for all observed nonlinearities. Second, the response up to a
frequency of about 1MHz can be well fitted assuming the sum of two
single-poled functions with cut-off frequencies of $1.6$ $%
\operatorname{kHz}%
$ and $119$ $%
\operatorname{kHz}%
$. It has been attributed to the response related to convective and conductive
heat exchange of the cavity mode with its environment
\cite{BraginskyGorodetsky1989,SchliesserDelHaye2006}. Concerning this
thermal-related response function, it is important to note that a temperature
change to the silica microcavity causes resonant frequency shifts via both a
change in the refractive index and a displacement due to thermo-mechanical
expansion. However, temperature-induced index changes dominate over thermal
expansion by a factor of at least 15 for glass. Therefore, the
thermo-mechanical contribution to the low-frequency-response shown in Figure
\ref{Figure18} is, at least, one order-of-magnitude smaller than the total
thermal response. \bigskip

Finally, at $58%
\operatorname{MHz}%
$ in the response spectrum, the radiation-pressure-driven mechanical response
is observable, which (inset Fig. 22) is a factor of $\times260$ stronger than
the Kerr nonlinearity in the present device. It is emphasized that this ratio
is in quantitative agreement with the theoretically predicted
Kerr-to-radiation pressure ratio. From the well-identified frequency response
of the aforementioned thermal effects it is possible to conclude that the
thermal effect contribution to the interaction of the cavity field with the
$58%
\operatorname{MHz}%
$ radial breathing mode is at least 2 orders of magnitude too weak to explain
the observed effects. Consequently, the response measurements give unambiguous
proof that such thermal effects contribute less than 1 part in 100 to the
observed cooling (or amplification) rate.

At a basic level, it is of significant importance that \textit{optical forces}
are dominant. These forces are both conservative and broadband, allowing for a
Hamiltonian formulation\cite{Law1995}, features that are fundamental to
cavity-QOM physics as well as to potential applications of cavity-OM as a technology.

\subsection{Quantum limits of radiation pressure back-action cooling \qquad}
An important question is what limiting temperature is achievable with the
radiation pressure back-action cooling technique as described above. Indeed,
ground state cooling of harmonically bound ions and atoms has first been
demonstrated by Wineland\cite{DiedrichBergquist1989} almost two decades ago
and has led to a remarkable set of advances in Atomic Physics, such as the
generation of non-classical states of ion motion\cite{MeekhofMonroe1996} or
Schr\"{o}dinger cat states\cite{MonroeMeekhof1996}. Note that ground state
cooling of mechanical oscillators is particularly challenging as the
temperature corresponding to a single phonon corresponds to 50 $\mu$K for a 1
MHz oscillator, or 50 mK for a 1 GHz oscillator, which are only\ accessible by
dilution refridgerators. Radiation-pressure back-action cooling might provide
a route to achieve yet lower temperatures.

However, even the early work of Braginsky did not address the fundamental
question of whether ground-state cooling is possible using the described
mechanisms. Recently, two theoretical
papers\cite{MarquardtChen2007,WilsonRaeNooshi2007} have extended the classical
theory of radiation-pressure back-action cooling to the quantum regime and
shown the close relationship that cavity back-action cooling has with the
laser cooling of harmonically bound atoms and ions. While a detailed
description of the theoretical framework can be found in
Refs.\cite{MarquardtChen2007,WilsonRaeNooshi2007} we here briefly recapitulate
the main results in terms of the fundamental temperature limit.

In the quantum mechanical treatment, the mechanical oscillator is described in
terms of its average occupancy $n$, where $n=0$ designates the quantum ground
state (in which the harmonic oscillator energy only contains the zero-point
energy contribution). In brief and as shown in
Refs.\cite{MarquardtChen2007,WilsonRaeNooshi2007}, the quantum mechanical
cooling limit is due to the fact that cooling proceeds both by motional
increasing and decreasing processes. Motional decreasing (increasing)
processes occur with a rate $R^{BSB}\propto\eta^{2}A^{-}n$ $(R^{RSB}%
\propto\eta^{2}A^{+}(n+1))$ , where $n$ is the phonon occupancy, $A^{\pm}$
$\propto\frac{1}{4(\Delta\mp\Omega_{m})^{2}+\kappa^{2}}$are the lorentzian
weights of the Stokes and anti-Stokes sidebands and $\eta=$ $\frac{\omega
_{0}x_{0}}{\Omega_{m}R}$an effective parameter\cite{WilsonRaeNooshi2007} with
$x_{0}=\sqrt{\frac{\hslash}{m_{eff}\Omega_{m}}}$being the zero point motion of
the mechanical oscillator mode. Detailed balance\cite{LeibfriedBlatt2003} of
the motional increasing and decreasing processes then yields the minimum final
occupancy (neglecting reservoir heating):
\begin{equation}
n_{f}=\frac{A^{+}}{A^{-}-A^{+}}%
\end{equation}

In the unresolved $\kappa\gg\Omega_{m}$ side-band case, (which has been
referred to as \textquotedblleft weak binding\textquotedblright\ in Atomic
Laser cooling\cite{WinelandItano1979}) this prevents ground state cooling as:
\begin{equation}
n_{f}\approx\frac{\kappa}{4\Omega_{m}}\gg1
\end{equation}

Converting this expression into an equivalent temperature $T_{D}\approx
\frac{\hbar\kappa}{4k_{B}}$, yields the \textit{Doppler temperature} (as known
in Atomic Laser Cooling). On the other hand, occupancies well below unity can
be attained in the resolved sideband case $\Omega_{m}\,\gg\kappa$ yielding:
\begin{equation}
n_{f}\approx\frac{\kappa^{2}}{16\Omega_{m}^{2}}\ll1
\end{equation}

If also the contribution from reservoir heating is included, the final
occupancy that can be achieved then takes the form%

\[
n_{f}=\frac{A^{+}}{A^{-}-A^{+}}+\frac{\Gamma_{m}}{\Gamma_{m}+\Gamma}n_{R}%
\]

Here the $n_{R}$ average denotes the average occupancy of the harmonic
oscillator prior to applying radiation pressure cooling. As before, this limit
is valid over the range where the cooling satisfies $\frac{\Gamma_{m}+\Gamma
}{\Gamma_{m}}<n_{R\text{ }}$and $\Gamma<\kappa$. Note that from the above
expression it becomes clear that pre-cooling of the mechanical oscillators is
advantageous and is indeed presently undertaken by several
groups\cite{ThompsonZwickl2007}. Resolved sideband cooling has also recently
been demonstrated experimentally\cite{SchliesserRiviere2007}.

\subsection{Physical Interpretation of the Quantum limits of Back-action
Cooling}

We next give some physical intuition into the above limits. It is worth noting
that the above introduced temperature limits are in fact a direct
manifestation of the \textit{Heisenberg uncertainty principle} and are of
rather general validity. This can be understood as follows. As shown in
Section 2 the mechanical motion produces blue-shifted sidebands which remove
mechanical quanta from the mechanical oscillator. These generated photons
decay during the cavity lifetime $1/\kappa$ into the outside world (i.e. into
the tapered optical fiber waveguide, or through the finite reflection of the
mirror in the case of a Fabry-Perot). The finite decay time of the cavity
however entails that the energy that is carried away by a blue-shifted photon
has a Heisenberg limited uncertainty of $\Delta E=\hslash/\Delta
t\approx\hslash\kappa$. Consequently, the mean energy of the mechanical
oscillator $\ E=\hslash\Omega_{m}(n+\frac{1}{2})\,~$cannot be lower than this
limit, implying a final Doppler temperature of $T_{D}\approx\hslash
\kappa/k_{B}$ in entire analogy to Laser Cooling in Atomic
Physics\cite{Stenholm1986}. This Doppler temperature entails that the final
occupancy is much larger than unity in the case where the mechanical
oscillator frequency is smaller than the cavity line-width (since $\kappa
\gg\Omega_{m}$). In contrast, ground state cooling is possible when then
energy quantum of the mechanical oscillator ($\hslash\Omega_{m}$) is made
large in comparison with the energy scale set by the Doppler temperature. This
required the \textquotedblleft strong binding condition\textquotedblright\ to
be satisfied, i.e. $\Omega_{m}\gg\kappa$.

A second consideration that can be used to successfully estimate the quantum
limits of backaction cooling is to consider the mechanical mirror as
performing a measurement on the intra-cavity photons. As in the case of
photo-detector shot noise, the random arrival time of photons onto the mirror
will entail a fluctuation of the radiation pressure force that will heat the
mechanical oscillator. If $N$ is the average number of photons in the cavity,
the fluctuation of photon numbers (governed by Poissonian statistics) is given
by $\sqrt{N}$, entailing a radiation pressure force fluctuation of
$\left\langle \Delta F_{RP}^{2}\right\rangle ^{1/2}=\sqrt{N}\frac
{2\hslash\kappa}{T_{rt}}$. Approximating this fluctuation as a white noise
spectrum over the cavity bandwidth, the corresponding force spectral density
of the radiation pressure quantum noise is given by $\delta F_{RP}^{2}%
(\Omega)=\left\langle \Delta F_{RP}^{2}\right\rangle /\kappa.$ This
effectively white spectrum provides a driving force for the mechanical
oscillator, and can be used to determine the final temperature via the
relation $k_{B}T_{f}=\int\Omega^{2}m_{eff}\left\vert \chi_{eff}(\Omega)\delta
F_{RP}(\Omega)\right\vert ^{2}d\Omega=\Gamma^{-1}m_{eff}\left\vert \delta
F_{RP}(\Omega_{m})\right\vert ^{2}$ (where $\chi_{eff}(\Omega)$ is the
modified mechanical oscillators susceptibility). Two limits can be derived.
First, assuming that the cavity line-width exceeds the mechanical oscillator
frequency, the noise spectrum is approximately given by $\delta F_{RP}%
^{2}(\Omega)\approx\frac{P}{\hbar\omega}\frac{1}{2}\tau(\frac{\hslash
k}{T_{rt}})^{2}=\frac{P\hslash\omega}{2L^{2}}\tau^{2}$ which yields i.e.
$k_{B}T_{f}\approx\hslash\kappa/4$ (i.e. $n_{f}\approx\frac{\kappa}%
{4\Omega_{m}}$ recovering the result derived in
\cite{MarquardtChen2007,WilsonRaeNooshi2007}). Here we have assumed a Fabry
Perot cavity of length $L$. On the other hand considering the resolved
sideband regime (i.e. assuming laser and $\Delta=-\Omega_{m}$) the spectral
density of the radiation pressure quantum fluctuations at the frequency of the
mechanical oscillator are given $\delta F_{RP}^{2}(\Omega_{m})\approx
\frac{P\hbar\omega}{2L^{2}\Omega_{m}}$ by yielding the result $k_{B}%
T_{f}=\frac{\hslash\Omega_{m}}{2}(1+\frac{\kappa^{2}}{8\Omega_{m}^{2}})$ or
$n_{f}\approx\frac{\kappa^{2}}{16\Omega_{m}^{2}}$. Thus, the final occupancy
in the unresolved and resolved sideband case can be understood as arising from
the quantum fluctuation of the intra-cavity field and are in agreement with
the results of a rigorous calculation. \

\section{Summary and Outlook}
In summary, we have described dynamical effects of radiation pressure in
Cavity Opto-Mechanics. Specifically, both amplification and cooling of
mechanical eigenmodes have been described as manifestations of
finite-cavity-delay on mechanical oscillator's dynamics. Beyond the described
phenomena of dynamical back-action, the Physics that can be studied in the
field of Cavity Optomechanics encompasses several other areas of
investigation. For instance, the classical dynamics of cavity Opto-mechanical
systems exhibits a wide range of phenomena, ranging from dynamical
multistability\cite{MarquardtHarris2006}, static
bistability\cite{DorselMcCullen1983} to chaotic regimes\cite{CarmonCross2007},
some of which have already been observed in an experimental setting and allow
one to create switchable tunable optical filters\cite{EichenfeldMichael2007}.
Furthermore, parametric amplification of mechanical modes constitutes an
entirely new way of creating the analogue of a `photonic quartz
oscillator\textquotedblright\ which is driven purely by the radiation pressure
of light and whose line-width is limited by thermal noise. With continued
improvements in mechanical Q that are already underway to address the
requirements of cooling-related-research, there is the potential for
realization of a new class of ultra-stable, narrow linewidth rf oscillators.

On the cooling side, there is a rich history of theoretical proposals
pertaining to entangling mechanical oscillators with a light field using
radiation
pressure\cite{ManciniTombesi1994,GiovannettiMancini2001,ManciniGiovannetti2002}%
. Most of these will require achieving temperatures at which the mechanical
system is close-to, or at, the quantum ground state. With the rapid progress
towards realization of ground-state cooling, it now seems likely that many of
these ideas and proposals can be tested over the next decade. Similarly,
macroscopic mechanical modes can provide a new medium in which to explore
quantum information phenomena, as has been true in the rich scientific arena
of cavity QED\cite{Kimble1998}. Indeed, one can view cavity QOM as the logical
extension of cavity QED into the macroscopic realm.

Another interesting goal is to reach a regime where cavity QOM phenomena
become observable. For example, a regime where the back-action from the
quantum noise of the radiation pressure dominates the thermal
noise\cite{JacobsTittonen1999, TittonenBreitenbach1999} would enable
experimental realizations of proposals such as the quantum non-demolition
measurement of photon number or pondermotive
squeezing\cite{ManciniTombesi1994}. Moreover, recent work from
Yale\cite{ThompsonZwickl2007} has demonstrated Cavity OM systems which realize
a quadratic coupling to the mechanical coordinate, thereby lending themselves
to perform QND measurements of the mechanical eigenstate.

As originally noted, both cooling\cite{BraginskyVyatchanin2002}
(\textquotedblleft tranquilizer\textquotedblright) and parametric
instability\cite{BraginskyStrigin2002} were first conceived theoretically in
the context of gravitational-wave detection by Braginsky. No doubt, the better
understanding of these phenomena gained by their demonstration in the
micro-scale will benefit this important field. Along these lines, since the
initial observation of radiation pressure parametric instability in
microtoroids\cite{CarmonRokhsari2005,KippenbergRokhsari2005,RokhsariKippenberg2005}
in 2005, the MIT gravitational-wave group has reported observation of
parametric instability\cite{CorbittOttaway2006}. Even more recently they have
also reported the cooling of gram scale mirror modes\cite{CorbittChen2007}.
These results bode well for further progress in this field.

From a practical point-of-view, the ability to achieve cooling and oscillation
of micromechanical modes on a semiconductor chip bodes well for realization of
new technologies that could leverage these new tools. Specifically,
miniaturization and integration of these functions with electronics and other
optical functions is already possible because of this microfabricated,
chip-based platform. Also significant is that radiation-pressure cooling
through dynamic back action, as already noted, is a highly targeted form of
cooling in which a selected mechanical mode(s) can be precisely defined to
receive the benefit of cooling, while other modes remain at elevated or even
at room temperature. Indeed, the first demonstrations of backaction cooling
were to temperatures in the range of 10 K, but featured mechanical structures
that were otherwise uncooled and at room temperature. As a result, this novel
form of cooling offers ultra-low temperature performance with relatively low
power requirements (milli-Watts) and without the need for cryogenics, vacuum
handling, or any of the other necessities of conventional refrigeration. This
feature, above all others, would seem to offer the greatest advantages in
terms of new technologies.

Finally, the range of phenomena that have been described here extend over the
entire electromagnetic spectrum. Indeed, the concept of dynamic back action
was conceived-of first in the microwave realm\cite{Braginsky1977}. It is also
important to note that there are electro-mechanical systems that provide an
analogous form of back action cooling\cite{NaikBuu2006,BrownBritton2007}.
These systems can potentially provide a means to achieve the quantum ground-state.

In summary Cavity (Quantum) Opto-Mechanics represents many of the concepts of
atomic and molecular physics, however embodied in an entirely different
macroscale system. It is currently experiencing rapid experimental and
theoretical success in various laboratories worldwide and offers entire new
inroads for basic science and potentially new technologies. It seems clear
that this field is entering an exciting period of experimental science.

\section{Acknowledgements}
KJV acknowledges the Caltech Lee Center and DARPA for supporting this work.
TJK acknowledges support via a Max Planck Independent Junior Research Group, a
Marie Curie Excellence Grant (MEXT-CT-2006-042842) and the Nanoscience
Initiative Munich (NIM). The authors kindly thank Albert Schliesser, Olivier
Arcizet, Jens Dobrindt and Mani Hossein-Zadeh for contributions to this review.

\bibliographystyle{apsrev}
\bibliography{OEBibliography}

\begin{thebibliography}{80}
\expandafter\ifx\csname natexlab\endcsname\relax\def\natexlab#1{#1}\fi
\expandafter\ifx\csname bibnamefont\endcsname\relax
  \def\bibnamefont#1{#1}\fi
\expandafter\ifx\csname bibfnamefont\endcsname\relax
  \def\bibfnamefont#1{#1}\fi
\expandafter\ifx\csname citenamefont\endcsname\relax
  \def\citenamefont#1{#1}\fi
\expandafter\ifx\csname url\endcsname\relax
  \def\url#1{\texttt{#1}}\fi
\expandafter\ifx\csname urlprefix\endcsname\relax\def\urlprefix{URL }\fi
\providecommand{\bibinfo}[2]{#2}
\providecommand{\eprint}[2][]{\url{#2}}

\bibitem[{\citenamefont{Vahala}(2003)}]{Vahala2003}
\bibinfo{author}{\bibfnamefont{K.~J.} \bibnamefont{Vahala}},
  \bibinfo{journal}{Nature} \textbf{\bibinfo{volume}{424}},
  \bibinfo{pages}{839} (\bibinfo{year}{2003}).

\bibitem[{\citenamefont{Craighead}(2000)}]{Craighead2000}
\bibinfo{author}{\bibfnamefont{H.~G.} \bibnamefont{Craighead}},
  \bibinfo{journal}{Science} \textbf{\bibinfo{volume}{290}},
  \bibinfo{pages}{1532} (\bibinfo{year}{2000}).

\bibitem[{\citenamefont{Hansch and Schawlow}(1975)}]{Hansch1975}
\bibinfo{author}{\bibfnamefont{T.~W.} \bibnamefont{Hansch}} \bibnamefont{and}
  \bibinfo{author}{\bibfnamefont{A.~L.} \bibnamefont{Schawlow}},
  \bibinfo{journal}{Optics Communications} \textbf{\bibinfo{volume}{13}},
  \bibinfo{pages}{68} (\bibinfo{year}{1975}).

\bibitem[{\citenamefont{Wineland et~al.}(1978)\citenamefont{Wineland,
  Drullinger, and Walls}}]{WinelandDrullinger1978}
\bibinfo{author}{\bibfnamefont{D.~J.} \bibnamefont{Wineland}},
  \bibinfo{author}{\bibfnamefont{R.~E.} \bibnamefont{Drullinger}},
  \bibnamefont{and} \bibinfo{author}{\bibfnamefont{F.~L.} \bibnamefont{Walls}},
  \bibinfo{journal}{Physical Review Letters} \textbf{\bibinfo{volume}{40}},
  \bibinfo{pages}{1639} (\bibinfo{year}{1978}).

\bibitem[{\citenamefont{Chu et~al.}(1985)\citenamefont{Chu, Hollberg,
  Bjorkholm, Cable, and Ashkin}}]{ChuHollberg1985}
\bibinfo{author}{\bibfnamefont{S.}~\bibnamefont{Chu}},
  \bibinfo{author}{\bibfnamefont{L.}~\bibnamefont{Hollberg}},
  \bibinfo{author}{\bibfnamefont{J.~E.} \bibnamefont{Bjorkholm}},
  \bibinfo{author}{\bibfnamefont{A.}~\bibnamefont{Cable}}, \bibnamefont{and}
  \bibinfo{author}{\bibfnamefont{A.}~\bibnamefont{Ashkin}},
  \bibinfo{journal}{Physical Review Letters} \textbf{\bibinfo{volume}{55}},
  \bibinfo{pages}{48} (\bibinfo{year}{1985}).

\bibitem[{\citenamefont{Stenholm}(1986)}]{Stenholm1986}
\bibinfo{author}{\bibfnamefont{S.}~\bibnamefont{Stenholm}},
  \bibinfo{journal}{Reviews of Modern Physics} \textbf{\bibinfo{volume}{58}},
  \bibinfo{pages}{699} (\bibinfo{year}{1986}).

\bibitem[{\citenamefont{Caves}(1981)}]{Caves1981}
\bibinfo{author}{\bibfnamefont{C.~M.} \bibnamefont{Caves}},
  \bibinfo{journal}{Physical Review D} \textbf{\bibinfo{volume}{23}},
  \bibinfo{pages}{1693} (\bibinfo{year}{1981}).

\bibitem[{\citenamefont{Jacobs et~al.}(1999)\citenamefont{Jacobs, Tittonen,
  Wiseman, and Schiller}}]{JacobsTittonen1999}
\bibinfo{author}{\bibfnamefont{K.}~\bibnamefont{Jacobs}},
  \bibinfo{author}{\bibfnamefont{I.}~\bibnamefont{Tittonen}},
  \bibinfo{author}{\bibfnamefont{H.~M.} \bibnamefont{Wiseman}},
  \bibnamefont{and} \bibinfo{author}{\bibfnamefont{S.}~\bibnamefont{Schiller}},
  \bibinfo{journal}{Physical Review A} \textbf{\bibinfo{volume}{60}},
  \bibinfo{pages}{538} (\bibinfo{year}{1999}).

\bibitem[{\citenamefont{Tittonen et~al.}(1999)\citenamefont{Tittonen,
  Breitenbach, Kalkbrenner, Muller, Conradt, Schiller, Steinsland, Blanc, and
  de~Rooij}}]{TittonenBreitenbach1999}
\bibinfo{author}{\bibfnamefont{I.}~\bibnamefont{Tittonen}},
  \bibinfo{author}{\bibfnamefont{G.}~\bibnamefont{Breitenbach}},
  \bibinfo{author}{\bibfnamefont{T.}~\bibnamefont{Kalkbrenner}},
  \bibinfo{author}{\bibfnamefont{T.}~\bibnamefont{Muller}},
  \bibinfo{author}{\bibfnamefont{R.}~\bibnamefont{Conradt}},
  \bibinfo{author}{\bibfnamefont{S.}~\bibnamefont{Schiller}},
  \bibinfo{author}{\bibfnamefont{E.}~\bibnamefont{Steinsland}},
  \bibinfo{author}{\bibfnamefont{N.}~\bibnamefont{Blanc}}, \bibnamefont{and}
  \bibinfo{author}{\bibfnamefont{N.~F.} \bibnamefont{de~Rooij}},
  \bibinfo{journal}{Physical Review A} \textbf{\bibinfo{volume}{59}},
  \bibinfo{pages}{1038} (\bibinfo{year}{1999}).

\bibitem[{\citenamefont{Braginsky}(1977)}]{Braginsky1977}
\bibinfo{author}{\bibfnamefont{V.~B.} \bibnamefont{Braginsky}},
  \emph{\bibinfo{title}{Measurement of Weak Forces in Physics Experiments}}
  (\bibinfo{publisher}{University of Chicago Press},
  \bibinfo{address}{Chicago}, \bibinfo{year}{1977}).

\bibitem[{\citenamefont{Braginsky and Khalili}(1992)}]{BraginskyKhalili1992}
\bibinfo{author}{\bibfnamefont{V.~B.} \bibnamefont{Braginsky}}
  \bibnamefont{and} \bibinfo{author}{\bibfnamefont{F.}~\bibnamefont{Khalili}},
  \emph{\bibinfo{title}{Quantum Measurement}} (\bibinfo{publisher}{Cambridge
  University Press}, \bibinfo{year}{1992}).

\bibitem[{\citenamefont{Mancini and Tombesi}(1994)}]{ManciniTombesi1994}
\bibinfo{author}{\bibfnamefont{S.}~\bibnamefont{Mancini}} \bibnamefont{and}
  \bibinfo{author}{\bibfnamefont{P.}~\bibnamefont{Tombesi}},
  \bibinfo{journal}{Physical Review A} \textbf{\bibinfo{volume}{49}},
  \bibinfo{pages}{4055} (\bibinfo{year}{1994}).

\bibitem[{\citenamefont{Bose et~al.}(1997)\citenamefont{Bose, Jacobs, and
  Knight}}]{BoseJacobs1997}
\bibinfo{author}{\bibfnamefont{S.}~\bibnamefont{Bose}},
  \bibinfo{author}{\bibfnamefont{K.}~\bibnamefont{Jacobs}}, \bibnamefont{and}
  \bibinfo{author}{\bibfnamefont{P.~L.} \bibnamefont{Knight}},
  \bibinfo{journal}{Physical Review A} \textbf{\bibinfo{volume}{56}},
  \bibinfo{pages}{4175} (\bibinfo{year}{1997}).

\bibitem[{\citenamefont{LaHaye et~al.}(2004)\citenamefont{LaHaye, Buu,
  Camarota, and Schwab}}]{LaHayeBuu2004}
\bibinfo{author}{\bibfnamefont{M.~D.} \bibnamefont{LaHaye}},
  \bibinfo{author}{\bibfnamefont{O.}~\bibnamefont{Buu}},
  \bibinfo{author}{\bibfnamefont{B.}~\bibnamefont{Camarota}}, \bibnamefont{and}
  \bibinfo{author}{\bibfnamefont{K.~C.} \bibnamefont{Schwab}},
  \bibinfo{journal}{Science} \textbf{\bibinfo{volume}{304}},
  \bibinfo{pages}{74} (\bibinfo{year}{2004}).

\bibitem[{\citenamefont{Naik et~al.}(2006)\citenamefont{Naik, Buu, LaHaye,
  Armour, Clerk, Blencowe, and Schwab}}]{NaikBuu2006}
\bibinfo{author}{\bibfnamefont{A.}~\bibnamefont{Naik}},
  \bibinfo{author}{\bibfnamefont{O.}~\bibnamefont{Buu}},
  \bibinfo{author}{\bibfnamefont{M.~D.} \bibnamefont{LaHaye}},
  \bibinfo{author}{\bibfnamefont{A.~D.} \bibnamefont{Armour}},
  \bibinfo{author}{\bibfnamefont{A.~A.} \bibnamefont{Clerk}},
  \bibinfo{author}{\bibfnamefont{M.~P.} \bibnamefont{Blencowe}},
  \bibnamefont{and} \bibinfo{author}{\bibfnamefont{K.~C.}
  \bibnamefont{Schwab}}, \bibinfo{journal}{Nature}
  \textbf{\bibinfo{volume}{443}}, \bibinfo{pages}{193} (\bibinfo{year}{2006}).

\bibitem[{\citenamefont{Brown et~al.}(2007)\citenamefont{Brown, Britton,
  Epstein, Chiaverini, Leibfried, and Wineland}}]{BrownBritton2007}
\bibinfo{author}{\bibfnamefont{K.}~\bibnamefont{Brown}},
  \bibinfo{author}{\bibfnamefont{J.}~\bibnamefont{Britton}},
  \bibinfo{author}{\bibfnamefont{R.}~\bibnamefont{Epstein}},
  \bibinfo{author}{\bibfnamefont{J.}~\bibnamefont{Chiaverini}},
  \bibinfo{author}{\bibfnamefont{D.}~\bibnamefont{Leibfried}},
  \bibnamefont{and} \bibinfo{author}{\bibfnamefont{D.}~\bibnamefont{Wineland}},
  \bibinfo{journal}{Physical Review Letters} \textbf{\bibinfo{volume}{99}},
  \bibinfo{pages}{137205} (\bibinfo{year}{2007}).

\bibitem[{\citenamefont{Dorsel et~al.}(1983)\citenamefont{Dorsel, McCullen,
  Meystre, Vignes, and Walther}}]{DorselMcCullen1983}
\bibinfo{author}{\bibfnamefont{A.}~\bibnamefont{Dorsel}},
  \bibinfo{author}{\bibfnamefont{J.~D.} \bibnamefont{McCullen}},
  \bibinfo{author}{\bibfnamefont{P.}~\bibnamefont{Meystre}},
  \bibinfo{author}{\bibfnamefont{E.}~\bibnamefont{Vignes}}, \bibnamefont{and}
  \bibinfo{author}{\bibfnamefont{H.}~\bibnamefont{Walther}},
  \bibinfo{journal}{Physical Review Letters} \textbf{\bibinfo{volume}{51}},
  \bibinfo{pages}{1550} (\bibinfo{year}{1983}).

\bibitem[{\citenamefont{Sheard et~al.}(2004)\citenamefont{Sheard, Gray,
  Mow-Lowry, McClelland, and Whitcomb}}]{SheardGray2004}
\bibinfo{author}{\bibfnamefont{B.~S.} \bibnamefont{Sheard}},
  \bibinfo{author}{\bibfnamefont{M.~B.} \bibnamefont{Gray}},
  \bibinfo{author}{\bibfnamefont{C.~M.} \bibnamefont{Mow-Lowry}},
  \bibinfo{author}{\bibfnamefont{D.~E.} \bibnamefont{McClelland}},
  \bibnamefont{and} \bibinfo{author}{\bibfnamefont{S.~E.}
  \bibnamefont{Whitcomb}}, \bibinfo{journal}{Physical Review A}
  \textbf{\bibinfo{volume}{69}} (\bibinfo{year}{2004}).

\bibitem[{\citenamefont{Braginsky et~al.}(2001)\citenamefont{Braginsky,
  Strigin, and Vyatchanin}}]{BraginskyStrigin2001}
\bibinfo{author}{\bibfnamefont{V.~B.} \bibnamefont{Braginsky}},
  \bibinfo{author}{\bibfnamefont{S.~E.} \bibnamefont{Strigin}},
  \bibnamefont{and} \bibinfo{author}{\bibfnamefont{S.~P.}
  \bibnamefont{Vyatchanin}}, \bibinfo{journal}{Physics Letters A}
  \textbf{\bibinfo{volume}{287}}, \bibinfo{pages}{331} (\bibinfo{year}{2001}).

\bibitem[{\citenamefont{Kippenberg et~al.}(2005)\citenamefont{Kippenberg,
  Rokhsari, Carmon, Scherer, and Vahala}}]{KippenbergRokhsari2005}
\bibinfo{author}{\bibfnamefont{T.~J.} \bibnamefont{Kippenberg}},
  \bibinfo{author}{\bibfnamefont{H.}~\bibnamefont{Rokhsari}},
  \bibinfo{author}{\bibfnamefont{T.}~\bibnamefont{Carmon}},
  \bibinfo{author}{\bibfnamefont{A.}~\bibnamefont{Scherer}}, \bibnamefont{and}
  \bibinfo{author}{\bibfnamefont{K.~J.} \bibnamefont{Vahala}},
  \bibinfo{journal}{Physical Review Letters} \textbf{\bibinfo{volume}{95}},
  \bibinfo{pages}{033901} (\bibinfo{year}{2005}).

\bibitem[{\citenamefont{Rokhsari et~al.}(2005)\citenamefont{Rokhsari,
  Kippenberg, Carmon, and Vahala}}]{RokhsariKippenberg2005}
\bibinfo{author}{\bibfnamefont{H.}~\bibnamefont{Rokhsari}},
  \bibinfo{author}{\bibfnamefont{T.~J.} \bibnamefont{Kippenberg}},
  \bibinfo{author}{\bibfnamefont{T.}~\bibnamefont{Carmon}}, \bibnamefont{and}
  \bibinfo{author}{\bibfnamefont{K.~J.} \bibnamefont{Vahala}},
  \bibinfo{journal}{Optics Express} \textbf{\bibinfo{volume}{13}},
  \bibinfo{pages}{5293} (\bibinfo{year}{2005}).

\bibitem[{\citenamefont{Carmon et~al.}(2005)\citenamefont{Carmon, Rokhsari,
  Yang, Kippenberg, and Vahala}}]{CarmonRokhsari2005}
\bibinfo{author}{\bibfnamefont{T.}~\bibnamefont{Carmon}},
  \bibinfo{author}{\bibfnamefont{H.}~\bibnamefont{Rokhsari}},
  \bibinfo{author}{\bibfnamefont{L.}~\bibnamefont{Yang}},
  \bibinfo{author}{\bibfnamefont{T.~J.} \bibnamefont{Kippenberg}},
  \bibnamefont{and} \bibinfo{author}{\bibfnamefont{K.~J.}
  \bibnamefont{Vahala}}, \bibinfo{journal}{Physical Review Letters}
  \textbf{\bibinfo{volume}{94}} (\bibinfo{year}{2005}).

\bibitem[{\citenamefont{Braginsky and
  Vyatchanin}(2002)}]{BraginskyVyatchanin2002}
\bibinfo{author}{\bibfnamefont{V.~B.} \bibnamefont{Braginsky}}
  \bibnamefont{and} \bibinfo{author}{\bibfnamefont{S.~P.}
  \bibnamefont{Vyatchanin}}, \bibinfo{journal}{Physics Letters A}
  \textbf{\bibinfo{volume}{293}}, \bibinfo{pages}{228} (\bibinfo{year}{2002}).

\bibitem[{\citenamefont{Arcizet
  et~al.}(2006{\natexlab{a}})\citenamefont{Arcizet, Cohadon, Briant, Pinard,
  and Heidmann}}]{ArcizetCohadon2006Nature}
\bibinfo{author}{\bibfnamefont{O.}~\bibnamefont{Arcizet}},
  \bibinfo{author}{\bibfnamefont{P.~F.} \bibnamefont{Cohadon}},
  \bibinfo{author}{\bibfnamefont{T.}~\bibnamefont{Briant}},
  \bibinfo{author}{\bibfnamefont{M.}~\bibnamefont{Pinard}}, \bibnamefont{and}
  \bibinfo{author}{\bibfnamefont{A.}~\bibnamefont{Heidmann}},
  \bibinfo{journal}{Nature} \textbf{\bibinfo{volume}{444}}, \bibinfo{pages}{71}
  (\bibinfo{year}{2006}{\natexlab{a}}).

\bibitem[{\citenamefont{Gigan et~al.}(2006)\citenamefont{Gigan, Bohm,
  Paternostro, Blaser, Langer, Hertzberg, Schwab, Bauerle, Aspelmeyer, and
  Zeilinger}}]{GiganBohm2006}
\bibinfo{author}{\bibfnamefont{S.}~\bibnamefont{Gigan}},
  \bibinfo{author}{\bibfnamefont{H.~R.} \bibnamefont{Bohm}},
  \bibinfo{author}{\bibfnamefont{M.}~\bibnamefont{Paternostro}},
  \bibinfo{author}{\bibfnamefont{F.}~\bibnamefont{Blaser}},
  \bibinfo{author}{\bibfnamefont{G.}~\bibnamefont{Langer}},
  \bibinfo{author}{\bibfnamefont{J.~B.} \bibnamefont{Hertzberg}},
  \bibinfo{author}{\bibfnamefont{K.~C.} \bibnamefont{Schwab}},
  \bibinfo{author}{\bibfnamefont{D.}~\bibnamefont{Bauerle}},
  \bibinfo{author}{\bibfnamefont{M.}~\bibnamefont{Aspelmeyer}},
  \bibnamefont{and}
  \bibinfo{author}{\bibfnamefont{A.}~\bibnamefont{Zeilinger}},
  \bibinfo{journal}{Nature} \textbf{\bibinfo{volume}{444}}, \bibinfo{pages}{67}
  (\bibinfo{year}{2006}).

\bibitem[{\citenamefont{Poggio et~al.}(2007)\citenamefont{Poggio, Degen, Mamin,
  and Rugar}}]{PoggioDegen2007}
\bibinfo{author}{\bibfnamefont{M.}~\bibnamefont{Poggio}},
  \bibinfo{author}{\bibfnamefont{C.~L.} \bibnamefont{Degen}},
  \bibinfo{author}{\bibfnamefont{H.~J.} \bibnamefont{Mamin}}, \bibnamefont{and}
  \bibinfo{author}{\bibfnamefont{D.}~\bibnamefont{Rugar}},
  \bibinfo{journal}{Physical Review Letters} \textbf{\bibinfo{volume}{99}}
  (\bibinfo{year}{2007}).

\bibitem[{\citenamefont{Schliesser et~al.}(2006)\citenamefont{Schliesser,
  Del'Haye, Nooshi, Vahala, and Kippenberg}}]{SchliesserDelHaye2006}
\bibinfo{author}{\bibfnamefont{A.}~\bibnamefont{Schliesser}},
  \bibinfo{author}{\bibfnamefont{P.}~\bibnamefont{Del'Haye}},
  \bibinfo{author}{\bibfnamefont{N.}~\bibnamefont{Nooshi}},
  \bibinfo{author}{\bibfnamefont{K.~J.} \bibnamefont{Vahala}},
  \bibnamefont{and} \bibinfo{author}{\bibfnamefont{T.~J.}
  \bibnamefont{Kippenberg}}, \bibinfo{journal}{Physical Review Letters}
  \textbf{\bibinfo{volume}{97}}, \bibinfo{pages}{243905}
  (\bibinfo{year}{2006}).

\bibitem[{\citenamefont{Cohadon et~al.}(1999)\citenamefont{Cohadon, Heidmann,
  and Pinard}}]{CohadonHeidmann1999}
\bibinfo{author}{\bibfnamefont{P.~F.} \bibnamefont{Cohadon}},
  \bibinfo{author}{\bibfnamefont{A.}~\bibnamefont{Heidmann}}, \bibnamefont{and}
  \bibinfo{author}{\bibfnamefont{M.}~\bibnamefont{Pinard}},
  \bibinfo{journal}{Physical Review Letters} \textbf{\bibinfo{volume}{83}},
  \bibinfo{pages}{3174} (\bibinfo{year}{1999}).

\bibitem[{\citenamefont{Corbitt et~al.}(2007)\citenamefont{Corbitt, Chen,
  Innerhofer, Muller-Ebhardt, Ottaway, Rehbein, Sigg, Whitcomb, Wipf, and
  Mavalvala}}]{CorbittChen2007}
\bibinfo{author}{\bibfnamefont{T.}~\bibnamefont{Corbitt}},
  \bibinfo{author}{\bibfnamefont{Y.~B.} \bibnamefont{Chen}},
  \bibinfo{author}{\bibfnamefont{E.}~\bibnamefont{Innerhofer}},
  \bibinfo{author}{\bibfnamefont{H.}~\bibnamefont{Muller-Ebhardt}},
  \bibinfo{author}{\bibfnamefont{D.}~\bibnamefont{Ottaway}},
  \bibinfo{author}{\bibfnamefont{H.}~\bibnamefont{Rehbein}},
  \bibinfo{author}{\bibfnamefont{D.}~\bibnamefont{Sigg}},
  \bibinfo{author}{\bibfnamefont{S.}~\bibnamefont{Whitcomb}},
  \bibinfo{author}{\bibfnamefont{C.}~\bibnamefont{Wipf}}, \bibnamefont{and}
  \bibinfo{author}{\bibfnamefont{N.}~\bibnamefont{Mavalvala}},
  \bibinfo{journal}{Physical Review Letters} \textbf{\bibinfo{volume}{98}},
  \bibinfo{pages}{150802} (\bibinfo{year}{2007}).

\bibitem[{\citenamefont{Thompson et~al.}(2007)\citenamefont{Thompson, Zwickl,
  Yarich, Marquardt, Girvin, and Harris}}]{ThompsonZwickl2007}
\bibinfo{author}{\bibfnamefont{J.~D.} \bibnamefont{Thompson}},
  \bibinfo{author}{\bibfnamefont{B.~M.} \bibnamefont{Zwickl}},
  \bibinfo{author}{\bibfnamefont{A.~M.} \bibnamefont{Yarich}},
  \bibinfo{author}{\bibfnamefont{F.}~\bibnamefont{Marquardt}},
  \bibinfo{author}{\bibfnamefont{S.~M.} \bibnamefont{Girvin}},
  \bibnamefont{and} \bibinfo{author}{\bibfnamefont{J.}~\bibnamefont{Harris}},
  \bibinfo{journal}{arXiv:0707.1724}  (\bibinfo{year}{2007}).

\bibitem[{\citenamefont{Mancini et~al.}(1998)\citenamefont{Mancini, Vitali, and
  Tombesi}}]{ManciniVitali1998}
\bibinfo{author}{\bibfnamefont{S.}~\bibnamefont{Mancini}},
  \bibinfo{author}{\bibfnamefont{D.}~\bibnamefont{Vitali}}, \bibnamefont{and}
  \bibinfo{author}{\bibfnamefont{P.}~\bibnamefont{Tombesi}},
  \bibinfo{journal}{Physical Review Letters} \textbf{\bibinfo{volume}{80}},
  \bibinfo{pages}{688} (\bibinfo{year}{1998}).

\bibitem[{\citenamefont{Vandermeer}(1985)}]{Vandermeer1985}
\bibinfo{author}{\bibfnamefont{S.}~\bibnamefont{Vandermeer}},
  \bibinfo{journal}{Reviews of Modern Physics} \textbf{\bibinfo{volume}{57}},
  \bibinfo{pages}{689} (\bibinfo{year}{1985}).

\bibitem[{\citenamefont{Kleckner and
  Bouwmeester}(2006)}]{KlecknerBouwmeester2006}
\bibinfo{author}{\bibfnamefont{D.}~\bibnamefont{Kleckner}} \bibnamefont{and}
  \bibinfo{author}{\bibfnamefont{D.}~\bibnamefont{Bouwmeester}},
  \bibinfo{journal}{Nature} \textbf{\bibinfo{volume}{444}}, \bibinfo{pages}{75}
  (\bibinfo{year}{2006}).

\bibitem[{\citenamefont{Arcizet
  et~al.}(2006{\natexlab{b}})\citenamefont{Arcizet, Cohadon, Briant, Pinard,
  Heidmann, Mackowski, Michel, Pinard, Francais, and
  Rousseau}}]{ArcizetCohadon2006}
\bibinfo{author}{\bibfnamefont{O.}~\bibnamefont{Arcizet}},
  \bibinfo{author}{\bibfnamefont{P.~F.} \bibnamefont{Cohadon}},
  \bibinfo{author}{\bibfnamefont{T.}~\bibnamefont{Briant}},
  \bibinfo{author}{\bibfnamefont{M.}~\bibnamefont{Pinard}},
  \bibinfo{author}{\bibfnamefont{A.}~\bibnamefont{Heidmann}},
  \bibinfo{author}{\bibfnamefont{J.~M.} \bibnamefont{Mackowski}},
  \bibinfo{author}{\bibfnamefont{C.}~\bibnamefont{Michel}},
  \bibinfo{author}{\bibfnamefont{L.}~\bibnamefont{Pinard}},
  \bibinfo{author}{\bibfnamefont{O.}~\bibnamefont{Francais}}, \bibnamefont{and}
  \bibinfo{author}{\bibfnamefont{L.}~\bibnamefont{Rousseau}},
  \bibinfo{journal}{Physical Review Letters} \textbf{\bibinfo{volume}{97}},
  \bibinfo{pages}{133601} (\bibinfo{year}{2006}{\natexlab{b}}).

\bibitem[{\citenamefont{Giovannetti et~al.}(2001)\citenamefont{Giovannetti,
  Mancini, and Tombesi}}]{GiovannettiMancini2001}
\bibinfo{author}{\bibfnamefont{V.}~\bibnamefont{Giovannetti}},
  \bibinfo{author}{\bibfnamefont{S.}~\bibnamefont{Mancini}}, \bibnamefont{and}
  \bibinfo{author}{\bibfnamefont{P.}~\bibnamefont{Tombesi}},
  \bibinfo{journal}{Europhysics Letters} \textbf{\bibinfo{volume}{54}},
  \bibinfo{pages}{559} (\bibinfo{year}{2001}).

\bibitem[{\citenamefont{Mancini et~al.}(2002)\citenamefont{Mancini,
  Giovannetti, Vitali, and Tombesi}}]{ManciniGiovannetti2002}
\bibinfo{author}{\bibfnamefont{S.}~\bibnamefont{Mancini}},
  \bibinfo{author}{\bibfnamefont{V.}~\bibnamefont{Giovannetti}},
  \bibinfo{author}{\bibfnamefont{D.}~\bibnamefont{Vitali}}, \bibnamefont{and}
  \bibinfo{author}{\bibfnamefont{P.}~\bibnamefont{Tombesi}},
  \bibinfo{journal}{Physical Review Letters} \textbf{\bibinfo{volume}{88}},
  \bibinfo{pages}{120401} (\bibinfo{year}{2002}).

\bibitem[{\citenamefont{Courty et~al.}(2003)\citenamefont{Courty, Heidmann, and
  Pinard}}]{CourtyHeidmann2003}
\bibinfo{author}{\bibfnamefont{J.~M.} \bibnamefont{Courty}},
  \bibinfo{author}{\bibfnamefont{A.}~\bibnamefont{Heidmann}}, \bibnamefont{and}
  \bibinfo{author}{\bibfnamefont{M.}~\bibnamefont{Pinard}},
  \bibinfo{journal}{Physical Review Letters} \textbf{\bibinfo{volume}{90}}
  (\bibinfo{year}{2003}).

\bibitem[{\citenamefont{Arcizet
  et~al.}(2006{\natexlab{c}})\citenamefont{Arcizet, Briant, Heidmann, and
  Pinard}}]{ArcizetBriant2006}
\bibinfo{author}{\bibfnamefont{O.}~\bibnamefont{Arcizet}},
  \bibinfo{author}{\bibfnamefont{T.}~\bibnamefont{Briant}},
  \bibinfo{author}{\bibfnamefont{A.}~\bibnamefont{Heidmann}}, \bibnamefont{and}
  \bibinfo{author}{\bibfnamefont{M.}~\bibnamefont{Pinard}},
  \bibinfo{journal}{Physical Review A} \textbf{\bibinfo{volume}{73}}
  (\bibinfo{year}{2006}{\natexlab{c}}).

\bibitem[{\citenamefont{Marshall et~al.}(2003)\citenamefont{Marshall, Simon,
  Penrose, and Bouwmeester}}]{MarshallSimon2003}
\bibinfo{author}{\bibfnamefont{W.}~\bibnamefont{Marshall}},
  \bibinfo{author}{\bibfnamefont{C.}~\bibnamefont{Simon}},
  \bibinfo{author}{\bibfnamefont{R.}~\bibnamefont{Penrose}}, \bibnamefont{and}
  \bibinfo{author}{\bibfnamefont{D.}~\bibnamefont{Bouwmeester}},
  \bibinfo{journal}{Physical Review Letters} \textbf{\bibinfo{volume}{91}}
  (\bibinfo{year}{2003}).

\bibitem[{\citenamefont{Hossein-Zadeh and Vahala}()}]{HosseinZadehVahala2007_2}
\bibinfo{author}{\bibfnamefont{M.}~\bibnamefont{Hossein-Zadeh}}
  \bibnamefont{and} \bibinfo{author}{\bibfnamefont{K.~J.} \bibnamefont{Vahala}}
  (????).

\bibitem[{\citenamefont{Povinelli et~al.}(2005)\citenamefont{Povinelli,
  Johnson, Loncar, Ibanescu, Smythe, Capasso, and
  Joannopoulos}}]{PovinelliJohnson2005}
\bibinfo{author}{\bibfnamefont{M.~L.} \bibnamefont{Povinelli}},
  \bibinfo{author}{\bibfnamefont{J.~M.} \bibnamefont{Johnson}},
  \bibinfo{author}{\bibfnamefont{M.}~\bibnamefont{Loncar}},
  \bibinfo{author}{\bibfnamefont{M.}~\bibnamefont{Ibanescu}},
  \bibinfo{author}{\bibfnamefont{E.~J.} \bibnamefont{Smythe}},
  \bibinfo{author}{\bibfnamefont{F.}~\bibnamefont{Capasso}}, \bibnamefont{and}
  \bibinfo{author}{\bibfnamefont{J.~D.} \bibnamefont{Joannopoulos}},
  \bibinfo{journal}{Optics Express} \textbf{\bibinfo{volume}{13}},
  \bibinfo{pages}{8286} (\bibinfo{year}{2005}).

\bibitem[{\citenamefont{Eichenfeld et~al.}(2007)\citenamefont{Eichenfeld,
  Michael, Perahia, and Painter}}]{EichenfeldMichael2007}
\bibinfo{author}{\bibfnamefont{M.}~\bibnamefont{Eichenfeld}},
  \bibinfo{author}{\bibfnamefont{C.}~\bibnamefont{Michael}},
  \bibinfo{author}{\bibfnamefont{R.}~\bibnamefont{Perahia}}, \bibnamefont{and}
  \bibinfo{author}{\bibfnamefont{O.}~\bibnamefont{Painter}},
  \bibinfo{journal}{Nature Photonics} \textbf{\bibinfo{volume}{1}},
  \bibinfo{pages}{416} (\bibinfo{year}{2007}).

\bibitem[{\citenamefont{Schwab and Roukes}(2005)}]{SchwabRoukes2005}
\bibinfo{author}{\bibfnamefont{K.~C.} \bibnamefont{Schwab}} \bibnamefont{and}
  \bibinfo{author}{\bibfnamefont{M.~L.} \bibnamefont{Roukes}},
  \bibinfo{journal}{Physics Today} \textbf{\bibinfo{volume}{58}},
  \bibinfo{pages}{36} (\bibinfo{year}{2005}).

\bibitem[{\citenamefont{Akahane et~al.}(2003)\citenamefont{Akahane, Asano,
  Song, and Noda}}]{AkahaneAsano2003}
\bibinfo{author}{\bibfnamefont{Y.}~\bibnamefont{Akahane}},
  \bibinfo{author}{\bibfnamefont{T.}~\bibnamefont{Asano}},
  \bibinfo{author}{\bibfnamefont{B.~S.} \bibnamefont{Song}}, \bibnamefont{and}
  \bibinfo{author}{\bibfnamefont{S.}~\bibnamefont{Noda}},
  \bibinfo{journal}{Nature} \textbf{\bibinfo{volume}{425}},
  \bibinfo{pages}{944} (\bibinfo{year}{2003}).

\bibitem[{\citenamefont{Schliesser et~al.}(2007)\citenamefont{Schliesser,
  Riviere, Anetsberger, Arcizet, and Kippenberg}}]{SchliesserRiviere2007}
\bibinfo{author}{\bibfnamefont{A.}~\bibnamefont{Schliesser}},
  \bibinfo{author}{\bibfnamefont{R.}~\bibnamefont{Riviere}},
  \bibinfo{author}{\bibfnamefont{G.}~\bibnamefont{Anetsberger}},
  \bibinfo{author}{\bibfnamefont{O.}~\bibnamefont{Arcizet}}, \bibnamefont{and}
  \bibinfo{author}{\bibfnamefont{I.~J.} \bibnamefont{Kippenberg}},
  \bibinfo{journal}{http://arxiv.org/abs/0709.4036}  (\bibinfo{year}{2007}).

\bibitem[{\citenamefont{Marquardt et~al.}(2007)\citenamefont{Marquardt, Chen,
  Clerk, and Girvin}}]{MarquardtChen2007}
\bibinfo{author}{\bibfnamefont{F.}~\bibnamefont{Marquardt}},
  \bibinfo{author}{\bibfnamefont{J.~P.} \bibnamefont{Chen}},
  \bibinfo{author}{\bibfnamefont{A.~A.} \bibnamefont{Clerk}}, \bibnamefont{and}
  \bibinfo{author}{\bibfnamefont{S.~M.} \bibnamefont{Girvin}},
  \bibinfo{journal}{Physical Review Letters} \textbf{\bibinfo{volume}{99}},
  \bibinfo{pages}{093902} (\bibinfo{year}{2007}).

\bibitem[{\citenamefont{Wilson-Rae et~al.}(2007)\citenamefont{Wilson-Rae,
  Nooshi, Zwerger, and Kippenberg}}]{WilsonRaeNooshi2007}
\bibinfo{author}{\bibfnamefont{I.}~\bibnamefont{Wilson-Rae}},
  \bibinfo{author}{\bibfnamefont{N.}~\bibnamefont{Nooshi}},
  \bibinfo{author}{\bibfnamefont{W.}~\bibnamefont{Zwerger}}, \bibnamefont{and}
  \bibinfo{author}{\bibfnamefont{T.~J.} \bibnamefont{Kippenberg}},
  \bibinfo{journal}{Physical Review Letters} \textbf{\bibinfo{volume}{99}},
  \bibinfo{pages}{093902} (\bibinfo{year}{2007}).

\bibitem[{\citenamefont{Law}(1995)}]{Law1995}
\bibinfo{author}{\bibfnamefont{C.~K.} \bibnamefont{Law}},
  \bibinfo{journal}{Physical Review A} \textbf{\bibinfo{volume}{51}},
  \bibinfo{pages}{2537} (\bibinfo{year}{1995}).

\bibitem[{\citenamefont{Haus}(1989)}]{Haus1989}
\bibinfo{author}{\bibfnamefont{H.~A.} \bibnamefont{Haus}},
  \emph{\bibinfo{title}{Electromagnetic fields and energy}}
  (\bibinfo{publisher}{Prentice Hall}, \bibinfo{address}{Englewood Cliff},
  \bibinfo{year}{1989}).

\bibitem[{\citenamefont{Pinard et~al.}(1999)\citenamefont{Pinard, Hadjar, and
  Heidmann}}]{PinardHadjar1999}
\bibinfo{author}{\bibfnamefont{M.}~\bibnamefont{Pinard}},
  \bibinfo{author}{\bibfnamefont{Y.}~\bibnamefont{Hadjar}}, \bibnamefont{and}
  \bibinfo{author}{\bibfnamefont{A.}~\bibnamefont{Heidmann}},
  \bibinfo{journal}{European Physical Journal D} \textbf{\bibinfo{volume}{7}},
  \bibinfo{pages}{107} (\bibinfo{year}{1999}).

\bibitem[{\citenamefont{Marquardt et~al.}(2006)\citenamefont{Marquardt, Harris,
  and Girvin}}]{MarquardtHarris2006}
\bibinfo{author}{\bibfnamefont{F.}~\bibnamefont{Marquardt}},
  \bibinfo{author}{\bibfnamefont{J.}~\bibnamefont{Harris}}, \bibnamefont{and}
  \bibinfo{author}{\bibfnamefont{S.}~\bibnamefont{Girvin}},
  \bibinfo{journal}{Physical Review Letters} \textbf{\bibinfo{volume}{96}}
  (\bibinfo{year}{2006}).

\bibitem[{\citenamefont{Hossein-Zadeh and
  Vahala}(2007)}]{HosseinZadehVahala2007}
\bibinfo{author}{\bibfnamefont{M.}~\bibnamefont{Hossein-Zadeh}}
  \bibnamefont{and} \bibinfo{author}{\bibfnamefont{K.~J.}
  \bibnamefont{Vahala}}, \bibinfo{journal}{Optics Letters}
  \textbf{\bibinfo{volume}{32}}, \bibinfo{pages}{1611} (\bibinfo{year}{2007}).

\bibitem[{\citenamefont{Karrai}(2006)}]{Karrai2006}
\bibinfo{author}{\bibfnamefont{K.}~\bibnamefont{Karrai}},
  \bibinfo{journal}{Nature} \textbf{\bibinfo{volume}{444}}, \bibinfo{pages}{41}
  (\bibinfo{year}{2006}).

\bibitem[{\citenamefont{Vuletic and Chu}(2000)}]{VuleticChu2000}
\bibinfo{author}{\bibfnamefont{V.}~\bibnamefont{Vuletic}} \bibnamefont{and}
  \bibinfo{author}{\bibfnamefont{S.}~\bibnamefont{Chu}},
  \bibinfo{journal}{Physical Review Letters} \textbf{\bibinfo{volume}{84}},
  \bibinfo{pages}{3787} (\bibinfo{year}{2000}).

\bibitem[{\citenamefont{Maunz et~al.}(2004)\citenamefont{Maunz, Puppe,
  Schuster, Syassen, Pinkse, and Rempe}}]{MaunzPuppe2004}
\bibinfo{author}{\bibfnamefont{P.}~\bibnamefont{Maunz}},
  \bibinfo{author}{\bibfnamefont{T.}~\bibnamefont{Puppe}},
  \bibinfo{author}{\bibfnamefont{I.}~\bibnamefont{Schuster}},
  \bibinfo{author}{\bibfnamefont{N.}~\bibnamefont{Syassen}},
  \bibinfo{author}{\bibfnamefont{P.~W.~H.} \bibnamefont{Pinkse}},
  \bibnamefont{and} \bibinfo{author}{\bibfnamefont{G.}~\bibnamefont{Rempe}},
  \bibinfo{journal}{Nature} \textbf{\bibinfo{volume}{428}}, \bibinfo{pages}{50}
  (\bibinfo{year}{2004}).

\bibitem[{\citenamefont{Braginsky et~al.}(1989)\citenamefont{Braginsky,
  Gorodetsky, and Ilchenko}}]{BraginskyGorodetsky1989}
\bibinfo{author}{\bibfnamefont{V.~B.} \bibnamefont{Braginsky}},
  \bibinfo{author}{\bibfnamefont{M.~L.} \bibnamefont{Gorodetsky}},
  \bibnamefont{and} \bibinfo{author}{\bibfnamefont{V.~S.}
  \bibnamefont{Ilchenko}}, \bibinfo{journal}{Physics Letters A}
  \textbf{\bibinfo{volume}{137}}, \bibinfo{pages}{393} (\bibinfo{year}{1989}).

\bibitem[{\citenamefont{Kippenberg et~al.}(2003)\citenamefont{Kippenberg,
  Spillane, Armani, and Vahala}}]{KippenbergSpillane2003}
\bibinfo{author}{\bibfnamefont{T.~J.} \bibnamefont{Kippenberg}},
  \bibinfo{author}{\bibfnamefont{S.~M.} \bibnamefont{Spillane}},
  \bibinfo{author}{\bibfnamefont{D.~K.} \bibnamefont{Armani}},
  \bibnamefont{and} \bibinfo{author}{\bibfnamefont{K.~J.}
  \bibnamefont{Vahala}}, \bibinfo{journal}{Applied Physics Letters}
  \textbf{\bibinfo{volume}{83}}, \bibinfo{pages}{797} (\bibinfo{year}{2003}).

\bibitem[{\citenamefont{Armani et~al.}(2003)\citenamefont{Armani, Kippenberg,
  Spillane, and Vahala}}]{ArmaniKippenberg2003}
\bibinfo{author}{\bibfnamefont{D.~K.} \bibnamefont{Armani}},
  \bibinfo{author}{\bibfnamefont{T.~J.} \bibnamefont{Kippenberg}},
  \bibinfo{author}{\bibfnamefont{S.~M.} \bibnamefont{Spillane}},
  \bibnamefont{and} \bibinfo{author}{\bibfnamefont{K.~J.}
  \bibnamefont{Vahala}}, \bibinfo{journal}{Nature}
  \textbf{\bibinfo{volume}{421}}, \bibinfo{pages}{925} (\bibinfo{year}{2003}).

\bibitem[{\citenamefont{Ma et~al.}(2007)\citenamefont{Ma, Schliesser, Del'Haye,
  Dabirian, Anetsberger, and Kippenberg}}]{MaSchliesser2007}
\bibinfo{author}{\bibfnamefont{R.}~\bibnamefont{Ma}},
  \bibinfo{author}{\bibfnamefont{A.}~\bibnamefont{Schliesser}},
  \bibinfo{author}{\bibfnamefont{P.}~\bibnamefont{Del'Haye}},
  \bibinfo{author}{\bibfnamefont{A.}~\bibnamefont{Dabirian}},
  \bibinfo{author}{\bibfnamefont{G.}~\bibnamefont{Anetsberger}},
  \bibnamefont{and} \bibinfo{author}{\bibfnamefont{T.~J.}
  \bibnamefont{Kippenberg}}, \bibinfo{journal}{Optics Letters}
  \textbf{\bibinfo{volume}{32}}, \bibinfo{pages}{2200} (\bibinfo{year}{2007}).

\bibitem[{\citenamefont{Carmon et~al.}(2007)\citenamefont{Carmon, Cross, and
  Vahala}}]{CarmonCross2007}
\bibinfo{author}{\bibfnamefont{T.}~\bibnamefont{Carmon}},
  \bibinfo{author}{\bibfnamefont{M.~C.} \bibnamefont{Cross}}, \bibnamefont{and}
  \bibinfo{author}{\bibfnamefont{K.~J.} \bibnamefont{Vahala}},
  \bibinfo{journal}{Physical Review Letters} \textbf{\bibinfo{volume}{98}}
  (\bibinfo{year}{2007}), \bibinfo{note}{0031-9007}.

\bibitem[{\citenamefont{Carmon and Vahala}(2007)}]{CarmonVahala2007}
\bibinfo{author}{\bibfnamefont{T.}~\bibnamefont{Carmon}} \bibnamefont{and}
  \bibinfo{author}{\bibfnamefont{K.~J.} \bibnamefont{Vahala}},
  \bibinfo{journal}{Physical Review Letters} \textbf{\bibinfo{volume}{98}}
  (\bibinfo{year}{2007}).

\bibitem[{\citenamefont{Spillane et~al.}(2003)\citenamefont{Spillane,
  Kippenberg, Painter, and Vahala}}]{SpillaneKippenberg2003}
\bibinfo{author}{\bibfnamefont{S.~M.} \bibnamefont{Spillane}},
  \bibinfo{author}{\bibfnamefont{T.~J.} \bibnamefont{Kippenberg}},
  \bibinfo{author}{\bibfnamefont{O.~J.} \bibnamefont{Painter}},
  \bibnamefont{and} \bibinfo{author}{\bibfnamefont{K.~J.}
  \bibnamefont{Vahala}}, \bibinfo{journal}{Physical Review Letters}
  \textbf{\bibinfo{volume}{91}}, \bibinfo{pages}{art. no.}
  (\bibinfo{year}{2003}).

\bibitem[{\citenamefont{Spillane et~al.}(2002)\citenamefont{Spillane,
  Kippenberg, and Vahala}}]{SpillaneKippenberg2002}
\bibinfo{author}{\bibfnamefont{S.~M.} \bibnamefont{Spillane}},
  \bibinfo{author}{\bibfnamefont{T.~J.} \bibnamefont{Kippenberg}},
  \bibnamefont{and} \bibinfo{author}{\bibfnamefont{K.~J.}
  \bibnamefont{Vahala}}, \bibinfo{journal}{Nature}
  \textbf{\bibinfo{volume}{415}}, \bibinfo{pages}{621} (\bibinfo{year}{2002}).

\bibitem[{\citenamefont{Kippenberg et~al.}(2004)\citenamefont{Kippenberg,
  Spillane, and Vahala}}]{KippenbergSpillane2004}
\bibinfo{author}{\bibfnamefont{T.~J.} \bibnamefont{Kippenberg}},
  \bibinfo{author}{\bibfnamefont{S.~M.} \bibnamefont{Spillane}},
  \bibnamefont{and} \bibinfo{author}{\bibfnamefont{K.~J.}
  \bibnamefont{Vahala}}, \bibinfo{journal}{Physical Review Letters}
  \textbf{\bibinfo{volume}{93}} (\bibinfo{year}{2004}).

\bibitem[{\citenamefont{Hossein-Zadeh et~al.}(2006)\citenamefont{Hossein-Zadeh,
  Rokhsari, Hajimiri, and Vahala}}]{HosseinZadehRokhsari2006}
\bibinfo{author}{\bibfnamefont{M.}~\bibnamefont{Hossein-Zadeh}},
  \bibinfo{author}{\bibfnamefont{H.}~\bibnamefont{Rokhsari}},
  \bibinfo{author}{\bibfnamefont{A.}~\bibnamefont{Hajimiri}}, \bibnamefont{and}
  \bibinfo{author}{\bibfnamefont{K.~J.} \bibnamefont{Vahala}},
  \bibinfo{journal}{Physical Review A} \textbf{\bibinfo{volume}{74}}
  (\bibinfo{year}{2006}).

\bibitem[{\citenamefont{Rokhsari et~al.}(2006)\citenamefont{Rokhsari,
  Hossein-Zadeh, Hajimiri, and Vahala}}]{RokhsariHosseinZadeh2006}
\bibinfo{author}{\bibfnamefont{H.}~\bibnamefont{Rokhsari}},
  \bibinfo{author}{\bibfnamefont{M.}~\bibnamefont{Hossein-Zadeh}},
  \bibinfo{author}{\bibfnamefont{A.}~\bibnamefont{Hajimiri}}, \bibnamefont{and}
  \bibinfo{author}{\bibfnamefont{K.~J.} \bibnamefont{Vahala}},
  \bibinfo{journal}{Applied Physics Letters} \textbf{\bibinfo{volume}{89}}
  (\bibinfo{year}{2006}).

\bibitem[{\citenamefont{Metzger and Karrai}(2004)}]{MetzgerKarrai2004}
\bibinfo{author}{\bibfnamefont{C.~H.} \bibnamefont{Metzger}} \bibnamefont{and}
  \bibinfo{author}{\bibfnamefont{K.}~\bibnamefont{Karrai}},
  \bibinfo{journal}{Nature} \textbf{\bibinfo{volume}{432}},
  \bibinfo{pages}{1002} (\bibinfo{year}{2004}).

\bibitem[{\citenamefont{Carmon et~al.}(2004)\citenamefont{Carmon, Yang, and
  Vahala}}]{CarmonYang2004}
\bibinfo{author}{\bibfnamefont{T.}~\bibnamefont{Carmon}},
  \bibinfo{author}{\bibfnamefont{L.}~\bibnamefont{Yang}}, \bibnamefont{and}
  \bibinfo{author}{\bibfnamefont{K.~J.} \bibnamefont{Vahala}},
  \bibinfo{journal}{Optics Express} \textbf{\bibinfo{volume}{12}},
  \bibinfo{pages}{4742} (\bibinfo{year}{2004}).

\bibitem[{\citenamefont{Bjorklund et~al.}(1983)\citenamefont{Bjorklund,
  Levenson, Lenth, and Ortiz}}]{BjorklundLevenson1983}
\bibinfo{author}{\bibfnamefont{G.}~\bibnamefont{Bjorklund}},
  \bibinfo{author}{\bibfnamefont{M.}~\bibnamefont{Levenson}},
  \bibinfo{author}{\bibfnamefont{W.}~\bibnamefont{Lenth}}, \bibnamefont{and}
  \bibinfo{author}{\bibfnamefont{C.}~\bibnamefont{Ortiz}},
  \bibinfo{journal}{Applied Physics B Lasers and Optics}
  \textbf{\bibinfo{volume}{32}}, \bibinfo{pages}{145} (\bibinfo{year}{1983}).

\bibitem[{\citenamefont{Zalalutdinov et~al.}(2001)\citenamefont{Zalalutdinov,
  Zehnder, Olkhovets, Turner, Sekaric, Ilic, Czaplewski, Parpia, and
  Craighead}}]{ZalalutdinovZehnder2001}
\bibinfo{author}{\bibfnamefont{M.}~\bibnamefont{Zalalutdinov}},
  \bibinfo{author}{\bibfnamefont{A.}~\bibnamefont{Zehnder}},
  \bibinfo{author}{\bibfnamefont{A.}~\bibnamefont{Olkhovets}},
  \bibinfo{author}{\bibfnamefont{S.}~\bibnamefont{Turner}},
  \bibinfo{author}{\bibfnamefont{L.}~\bibnamefont{Sekaric}},
  \bibinfo{author}{\bibfnamefont{B.}~\bibnamefont{Ilic}},
  \bibinfo{author}{\bibfnamefont{D.}~\bibnamefont{Czaplewski}},
  \bibinfo{author}{\bibfnamefont{J.~M.} \bibnamefont{Parpia}},
  \bibnamefont{and} \bibinfo{author}{\bibfnamefont{H.~G.}
  \bibnamefont{Craighead}}, \bibinfo{journal}{Applied Physics Letters}
  \textbf{\bibinfo{volume}{79}}, \bibinfo{pages}{695} (\bibinfo{year}{2001}).

\bibitem[{\citenamefont{Treussart et~al.}(1998)\citenamefont{Treussart,
  Ilchenko, Roch, Hare, Lefevre-Seguin, Raimond, and
  Haroche}}]{TreussartIlchenko1998}
\bibinfo{author}{\bibfnamefont{F.}~\bibnamefont{Treussart}},
  \bibinfo{author}{\bibfnamefont{V.~S.} \bibnamefont{Ilchenko}},
  \bibinfo{author}{\bibfnamefont{J.~F.} \bibnamefont{Roch}},
  \bibinfo{author}{\bibfnamefont{J.}~\bibnamefont{Hare}},
  \bibinfo{author}{\bibfnamefont{V.}~\bibnamefont{Lefevre-Seguin}},
  \bibinfo{author}{\bibfnamefont{J.~M.} \bibnamefont{Raimond}},
  \bibnamefont{and} \bibinfo{author}{\bibfnamefont{S.}~\bibnamefont{Haroche}},
  \bibinfo{journal}{European Physical Journal D} \textbf{\bibinfo{volume}{1}},
  \bibinfo{pages}{235} (\bibinfo{year}{1998}).

\bibitem[{\citenamefont{Rokhsari and Vahala}(2005)}]{RokhsariVahala2005}
\bibinfo{author}{\bibfnamefont{H.}~\bibnamefont{Rokhsari}} \bibnamefont{and}
  \bibinfo{author}{\bibfnamefont{K.~J.} \bibnamefont{Vahala}},
  \bibinfo{journal}{Optics Letters} \textbf{\bibinfo{volume}{30}},
  \bibinfo{pages}{427} (\bibinfo{year}{2005}).

\bibitem[{\citenamefont{Diedrich et~al.}(1989)\citenamefont{Diedrich,
  Bergquist, Itano, and Wineland}}]{DiedrichBergquist1989}
\bibinfo{author}{\bibfnamefont{F.}~\bibnamefont{Diedrich}},
  \bibinfo{author}{\bibfnamefont{J.~C.} \bibnamefont{Bergquist}},
  \bibinfo{author}{\bibfnamefont{W.~M.} \bibnamefont{Itano}}, \bibnamefont{and}
  \bibinfo{author}{\bibfnamefont{D.~J.} \bibnamefont{Wineland}},
  \bibinfo{journal}{Physical Review Letters} \textbf{\bibinfo{volume}{62}},
  \bibinfo{pages}{403} (\bibinfo{year}{1989}).

\bibitem[{\citenamefont{Meekhof et~al.}(1996)\citenamefont{Meekhof, Monroe,
  King, Itano, and Wineland}}]{MeekhofMonroe1996}
\bibinfo{author}{\bibfnamefont{D.~M.} \bibnamefont{Meekhof}},
  \bibinfo{author}{\bibfnamefont{C.}~\bibnamefont{Monroe}},
  \bibinfo{author}{\bibfnamefont{B.~E.} \bibnamefont{King}},
  \bibinfo{author}{\bibfnamefont{W.~M.} \bibnamefont{Itano}}, \bibnamefont{and}
  \bibinfo{author}{\bibfnamefont{D.~J.} \bibnamefont{Wineland}},
  \bibinfo{journal}{Physical Review Letters} \textbf{\bibinfo{volume}{76}},
  \bibinfo{pages}{1796} (\bibinfo{year}{1996}).

\bibitem[{\citenamefont{Monroe et~al.}(1996)\citenamefont{Monroe, Meekhof,
  King, and Wineland}}]{MonroeMeekhof1996}
\bibinfo{author}{\bibfnamefont{C.}~\bibnamefont{Monroe}},
  \bibinfo{author}{\bibfnamefont{D.~M.} \bibnamefont{Meekhof}},
  \bibinfo{author}{\bibfnamefont{B.~E.} \bibnamefont{King}}, \bibnamefont{and}
  \bibinfo{author}{\bibfnamefont{D.~J.} \bibnamefont{Wineland}},
  \bibinfo{journal}{Science} \textbf{\bibinfo{volume}{272}},
  \bibinfo{pages}{1131} (\bibinfo{year}{1996}).

\bibitem[{\citenamefont{Leibfried et~al.}(2003)\citenamefont{Leibfried, Blatt,
  Monroe, and Wineland}}]{LeibfriedBlatt2003}
\bibinfo{author}{\bibfnamefont{D.}~\bibnamefont{Leibfried}},
  \bibinfo{author}{\bibfnamefont{R.}~\bibnamefont{Blatt}},
  \bibinfo{author}{\bibfnamefont{C.}~\bibnamefont{Monroe}}, \bibnamefont{and}
  \bibinfo{author}{\bibfnamefont{D.}~\bibnamefont{Wineland}},
  \bibinfo{journal}{Reviews of Modern Physics} \textbf{\bibinfo{volume}{75}},
  \bibinfo{pages}{281} (\bibinfo{year}{2003}).

\bibitem[{\citenamefont{Wineland and Itano}(1979)}]{WinelandItano1979}
\bibinfo{author}{\bibfnamefont{D.~J.} \bibnamefont{Wineland}} \bibnamefont{and}
  \bibinfo{author}{\bibfnamefont{W.~M.} \bibnamefont{Itano}},
  \bibinfo{journal}{Physical Review A} \textbf{\bibinfo{volume}{20}},
  \bibinfo{pages}{1521} (\bibinfo{year}{1979}).

\bibitem[{\citenamefont{Kimble}(1998)}]{Kimble1998}
\bibinfo{author}{\bibfnamefont{H.~J.} \bibnamefont{Kimble}},
  \bibinfo{journal}{Physica Scripta} \textbf{\bibinfo{volume}{T76}},
  \bibinfo{pages}{127} (\bibinfo{year}{1998}).

\bibitem[{\citenamefont{Braginsky et~al.}(2002)\citenamefont{Braginsky,
  Strigin, and Vyatchanin}}]{BraginskyStrigin2002}
\bibinfo{author}{\bibfnamefont{V.~B.} \bibnamefont{Braginsky}},
  \bibinfo{author}{\bibfnamefont{S.~E.} \bibnamefont{Strigin}},
  \bibnamefont{and} \bibinfo{author}{\bibfnamefont{S.~P.}
  \bibnamefont{Vyatchanin}}, \bibinfo{journal}{Physics Letters A}
  \textbf{\bibinfo{volume}{305}}, \bibinfo{pages}{111} (\bibinfo{year}{2002}).

\bibitem[{\citenamefont{Corbitt et~al.}(2006)\citenamefont{Corbitt, Ottaway,
  Innerhofer, Pelc, and Mavalvala}}]{CorbittOttaway2006}
\bibinfo{author}{\bibfnamefont{T.}~\bibnamefont{Corbitt}},
  \bibinfo{author}{\bibfnamefont{D.}~\bibnamefont{Ottaway}},
  \bibinfo{author}{\bibfnamefont{E.}~\bibnamefont{Innerhofer}},
  \bibinfo{author}{\bibfnamefont{J.}~\bibnamefont{Pelc}}, \bibnamefont{and}
  \bibinfo{author}{\bibfnamefont{N.}~\bibnamefont{Mavalvala}},
  \bibinfo{journal}{Physical Review A} \textbf{\bibinfo{volume}{74}}
  (\bibinfo{year}{2006}).

\end{thebibliography}

\end{document}